\documentclass[prd,aps,floatfix,showpacs,superscriptaddress]{revtex4}

\usepackage{graphicx}
\usepackage{amsmath}
\usepackage[psamsfonts]{amssymb}
\usepackage{longtable}

\newcommand{\be}{\begin{equation}}
\newcommand{\ee}{\end{equation}}
\newcommand{\bea}{\begin{eqnarray}}
\newcommand{\eea}{\end{eqnarray}}
\newcommand{\bi}{\begin{itemize}}
\newcommand{\ei}{\end{itemize}}

\def\simge{
    \mathrel{\rlap{\raise 0.511ex
        \hbox{$>$}}{\lower 0.511ex \hbox{$\sim$}}}}
\def\simle{
    \mathrel{\rlap{\raise 0.511ex
        \hbox{$<$}}{\lower 0.511ex \hbox{$\sim$}}}}

\begin{document}


\vspace*{-10mm}
\begin{flushright}
\normalsize RBRC-613
\end{flushright}

\title{Signatures of $S$-wave bound-state formation in finite volume}

\author{Shoichi Sasaki}
\email{ssasaki@phys.s.u-tokyo.ac.jp}
\affiliation{RIKEN BNL Research Center, Bldg.~510A, \\
Brookhaven National Laboratory, Upton, NY, 11973, U.S.A.}
\affiliation{Department of Physics, The University of Tokyo, \\
Hongo 7-3-1, Tokyo 113-0033, Japan}

\author{Takeshi Yamazaki}
\email{yamazaki@quark.phy.bnl.gov}
\altaffiliation[Present address: ]{University of Connecticut, Physics Department,
U-3046 2152 Hillside Road, Storrs, CT 06269-3046, USA}   
\affiliation{RIKEN BNL Research Center, Bldg.~510A, \\
Brookhaven National Laboratory, Upton, NY, 11973, U.S.A.}

\date{\today}

\begin{abstract}


We discuss formation of an $S$-wave bound state in finite volume
on the basis of L\"uscher's phase-shift formula.
It is found that although a bound-state pole condition 
is fulfilled only in the infinite volume limit, its modification by
the finite size corrections is exponentially suppressed
by the spatial extent $L$ in a finite box $L^3$. 
We also confirm that the appearance of the $S$-wave 
bound state is accompanied by an abrupt sign change of 
the $S$-wave scattering length even in finite volume
through numerical simulations. This distinctive behavior 
may help us to distinguish the loosely bound state from 
the lowest energy level of the scattering state in finite volume simulations.

\end{abstract}


\maketitle


 
\section{Introduction}
\label{Sec:1}
In the past few years, several new hadronic resonances have 
been discovered  in various experiments~\cite{Klempt:2004yz}. 
However, some of the states have unusual properties, 
which are not well understood from the viewpoint of the conventional 
quark-antiquark or three-quark states. It is a great challenge for
lattice QCD to answer the question, whether those states are really
exotic hadron states. 

We are especially interested in some candidates of hadronic molecular state:
the $\Lambda$(1405) resonance as an $\overline{K}N$ bound state,
the $f_0(980)$ and $a_0(980)$ resonances as $S$-wave bound states of 
$K\overline{K}$, the $X(3872)$ resonance as a weakly bound 
state of $D\overline{D}^*$, the $D_{s0}(2317)$ and $D_{s1}(2460)$ 
resonances as $D^{(*)}K$ bound states and so on~\cite{Rosner:2006vc}. 
Such states lie near and below their respective thresholds so that one can view them
as  ``loosely bound states" of two hadrons like a deuteron.

In the infinite volume, the loosely (near-threshold) bound state 
is well defined since there is no continuum
state below threshold. However, in a finite box on the lattice, all states have 
discrete energies. Even worse, the lowest energy level of the elastic scattering state
appears below threshold in the case if an interaction is attractive between
two particles~\cite{{Luscher:1985dn},{Luscher:1990ux}}. 
Therefore, there is an ambiguity to distinguish between the loosely bound state 
and the lowest scattering state in finite volume in this sense.

Signatures of bound-state formation in finite volume are
of main interest in this paper. 
We may begin with a naive question: what is the legitimate 
definition of the loosely bound state in the quantum mechanics? 
In the scattering theory~\cite{Newton:1982qc}, 
poles of the $S$-matrix or the scattering amplitude
correspond to bound states. It is also known that the appearance of the $S$-wave 
bound state is accompanied by an abrupt sign change of the $S$-wave scattering length~\cite{Newton:1982qc}.
It is interpreted that formation of one bound-state raises the phase shift 
at threshold by $\pi$. This particular feature is generalized as Levinson's 
theorem~\cite{Newton:1982qc}.
Thus, it is interesting to consider how the formation condition
of bound states is implemented in L\" uscher's finite size method, 
which is proposed as a general method for computing low-energy scattering 
phases of two particles in finite volume~\cite{{Luscher:1985dn},{Luscher:1990ux}}. 

In this paper, we discuss bound-state formation on the basis of the phase-shift formula 
in this method and then present our proposal for numerical simulations to 
distinguish the loosely bound state from the lowest scattering state in finite volume.
To exhibit the validity and efficiency of our proposal, we perform numerical studies
of the positronium spectroscopy in compact scalar QED model.
In the Higgs phase of $U(1)$ gauge dynamics, the photon is massive and then 
massive photons give rise to the short-ranged interparticle force between 
an electron and a positron exponentially damped. 
In this model, we can control positronium formation 
in variation with the strength of the interparticle force and then explore distinctive
signatures of the bound-state formation in finite volume. 

The organization of our paper is as follows. In Sec.~\ref{Sec:2}, 
we first give a brief review of L\" uscher's finite size 
method~\cite{{Luscher:1985dn},{Luscher:1990ux}} and discuss 
bound-state formation on the basis of the phase-shift formula in this method. 
Sec.~\ref{Sec:3} gives details of our utilized model, compact scalar QED, 
and its Monte Carlo simulations.
Secs.~\ref{Sec:4} and ~\ref{Sec:5} are devoted to discuss 
our numerical results in the $^1S_0$ and $^3S_1$ channels
of electron-positron system, respectively.
Finally, in Sec.~\ref{Sec:6}, we summarize the present work and
give our concluding remark. In addition, there are two appendices. 
In Appendices A, the sensitivity of mass spectra to choice of spatial 
boundary condition is discussed. We also demonstrate a specific volume 
dependence of the spectral amplitude for either the bound state or the 
lowest scattering state in Appendix B.

\section{Methodology}
\label{Sec:2}
\subsection{L\"uscher's finite size method for scattering phase shift}
\label{Sec:2-A}

Let us briefly review L\"uscher's finite size method~\cite{{Luscher:1985dn},{Luscher:1990ux}}. 
So far, several hadron scattering lengths, {\it e.g.}
$\pi$-$\pi$,  $\pi$-$K$, $\pi$-$N$, $K$-$N$, $N$-$N$ and $J/\psi$-hadron, 
have been successfully calculated 
by using this method~\cite{{Guagnelli:1990jb},{Sharpe:1992pp},{Gupta:1993rn},{Fukugita:1994ve},
{Aoki:2002in},{Aoki:2002ny},{Yamazaki:2004qb},{Aoki:2005uf},{Beane:2005rj},{Beane:2006mx},{Beane:2006gj},{Liu:2001ss},{Miao:2004gy},{Meng:2003gm},{Hasenfratz:2004qk},{Yokokawa:2006td}}.

The total energy of two-particle states in the center-of-mass frame is given by
%
%
\be
E_{AB}(p)=\sqrt{m_A^2+p^2}+\sqrt{m_B^2+p^2},
\ee
where $p$ is 
the relative momentum of two particles. 
In a finite box $L^3$ on the lattice, all momenta
are quantized and can be labeled by an integer $n$
as ${\bar p}_n$, which represents the $(n+1)$-th lowest momentum.
Therefore, all two-particle states have only discrete energies.

We introduce the scaled momentum as $q=L{\bar p}_n/2\pi$ 
with the spatial extent $L$ for periodic boundary condition. 
Although the value of $q^2$ takes an integer value in the non-interacting case, 
$q^2$ is no longer the integer due to the presence of the two-particle interaction. 
This particular feature can be observed through 
an energy shift relative to the energy of the non-interacting two particles,
%
%
\be
\Delta E = E_{AB}({\bar p}_n)-E_{AB}(p_n),
\ee
where the energy of non-interacting two-particle states
$E_{AB}(p_n)$ can be evaluated with
the quantized momentum $p_n$ in the free case
as $p_n=2\pi \sqrt{n}/L$ with an integer $n$. 

It has been shown by L\"uscher that this energy shift in a finite box 
with a spatial size $L$ can be translated into the $S$-wave phase shift $\delta_{0}$ 
through the relation~\cite{{Luscher:1985dn},{Luscher:1990ux}}:
%
%
\be
\tan \delta_{0}({\bar p}_n)=\frac{\pi^{3/2} \sqrt{q^2}}{{\cal Z}_{00}(1,q^2)}\;\;\;\;{\rm at}\;\;
q=L{\bar p}_n/2\pi , 
\label{Eq.LucsherFormula}
\ee
where the function ${\cal Z}_{00}(s,q^2)$ is an analytic continuation of
the generalized zeta function, 
${\cal Z}_{00}(s,q^2) \equiv \frac{1}{\sqrt{4\pi}}\sum_{{\bf n}\in Z^3}
({\bf n}^2 - q^2)^{-s}$, from the region $s>3/2$ to $s=1$.
The $S$-wave scattering length is defined through 
$a_0=\lim_{p\rightarrow 0}\tan \delta_0(p)/p$.

If the $S$-wave scattering length $a_0$ is sufficiently smaller than the spatial 
size $L$, one can make a Taylor expansion of the phase-shift formula
(\ref{Eq.LucsherFormula}) around $q^2=0$, and then obtain 
the asymptotic solution of Eq.~(\ref{Eq.LucsherFormula}).
Under the condition $p^2 \ll m_A^2$ and $m_B^2$, the solution is given  by 
%
%
\be
\Delta E_{q^2=0}
\approx-\frac{2\pi a_0}{\mu L^3}\left[
1+c_1 \frac{a_0}{L}+c_2 \left(\frac{a_0}{L}\right)^2
\right] +{\cal O}(L^{-6}),
\label{Eq.ScattL0}
\ee
which corresponds to the energy shift of the lowest ($n=0$) scattering state. 
The coefficients are $c_1=-2.837297$ and  
$c_2=6.375183$~\cite{{Luscher:1985dn},{Luscher:1990ux}}.
The reduced mass of two particles $\mu$ is given by $\mu=m_A\cdot m_B/(m_A+m_B)$. 
An important message is received from Eq.~(\ref{Eq.ScattL0}).
The lowest energy level of the elastic scattering state appears below threshold  
on the lattice if an interaction is weakly attractive ($a_0>0$) 
between two particles. This point makes it difficult to distinguish between
near-threshold bound states and scattering states on the lattice.

Here, it is worth noting that the large $L$ expansion formula~(\ref{Eq.ScattL0}) up to 
$O(L^{-4})$ gives no real solution of $a_0$ for the case 
$\Delta E < - \frac{\pi}{2|c_1|\mu L^2}$~\cite{Yokokawa:2006td}, 
while Eq.~(\ref{Eq.ScattL0}) with an expansion up to $O(L^{-4})$ and that 
up to $O(L^{-5})$ always possesses a real and negative solution of $a_0$ for $\Delta E>0$.  
A lower bound $\Delta E \ge - \frac{\pi}{2|c_1|\mu L^2}$ may be crucial to identify the 
observed state below threshold as the lowest energy level of the elastic scattering state.

For the second lowest ($n=1$) scattering state, we also obtain a different
asymptotic solution of Eq.~(\ref{Eq.LucsherFormula}), which is given 
by a Taylor expansion of the phase-shift formula (\ref{Eq.LucsherFormula}) around $q^2=1$ as
%
%
\be
\Delta E_{q^2=1}\approx 
-\frac{6\tan \delta_0({\bar p}_1)}{\mu L^2}\left[
1+c_1^{\prime} \tan \delta_0({\bar p}_1) + c_2^{\prime} \tan^2 \delta_0({\bar p}_1)\right]
+{\cal O}(L^{-6}),
\label{Eq.ScattL1}
\ee
where $c_1^{\prime}=-0.061367$ and $c_2^{\prime}=-0.354156$.
Although the sign of $\tan \delta_0$ is not uniquely related to the sign of the energy
shift, the resulting energy shift $\Delta E$ becomes positive (negative)
for the weak repulsive (attractive) interaction case ($|\delta_0| \simle 3\pi/5$).
Subsequently, one can derive the asymptotic solutions for the higher energy levels of the 
scattering state around $q^2=\nu \ge 2$ where $\nu={\bf n}^2$ for integer 3-dim vectors ${\bf n}\in Z^3$. 
For those asymptotic solutions, the corresponding relative momentum
${\bar p}_n$, which we will hereafter 
abbreviate as $p$, should vanish as $1/L$ with increasing $L$.

\subsection{Bound-state formation in L\"uscher's formula}
\label{Sec:2-B}

In quantum scattering theory, the formation condition of bound states is implemented 
as a pole in the $S$-matrix or scattering amplitude. Therefore, an important question 
naturally arises as to how bound-state formation is studied through 
L\"uscher's phase-shift formula~(\ref{Eq.LucsherFormula}) .

Intuitively, the pole condition of the $S$-matrix: $S={\rm e}^{2i\delta_0(p)} = \frac{\cot\delta_0(p)+i}{\cot\delta_0(p)-i}$ is expressed as
\be
\cot \delta_0(p) = i,
\ee
which is satisfied at $p^2=-\gamma^2$ where positive real $\gamma$ represents the binding momentum. In fact, as we will discuss in the following, such a condition is fulfilled 
only in the infinite volume. However the finite-volume corrections on this pole condition 
are exponentially suppressed by the size of spatial extent $L$.

For {\it negative} $q^2$, an exponentially convergent expression
of the zeta function ${\cal Z}_{00}(s, q^2)$ has been derived 
in Ref.~\cite{Elizalde:1997jv}.
For $s=1$, it is given by
%
%
\be
{\cal Z}_{00}(1,q^2)=-\pi^{3/2}\sqrt{-q^2}+\sum_{{\bf n}\in{\bf Z^3}}{}^{\prime}
\frac{\pi^{1/2}}{2\sqrt{\bf n^2}}e^{-2\pi \sqrt{-q^2{\bf n^2}}},
\label{Eq:ECSformula}
\ee
where $\sum^{\prime}_{{\bf n}\in Z^3}$ means the summation without ${\bf n}=(0,0,0)$.
We now insert Eq.~(\ref{Eq:ECSformula}) into Eq.~(\ref{Eq.LucsherFormula}) and then obtain
the following formula, which is mathematically equivalent to Eq.~(\ref{Eq.LucsherFormula})
for {\it negative} $q^2$:
%
%
\be
\cot \delta_0(p) = i + \frac{1}{2\pi i}\sum_{{\bf n}\in{Z^3}}{}^{\prime}
\frac{1}{\sqrt{-q^2{\bf n}^2}}e^{-2\pi\sqrt{-q^2{\bf n}^2}}.
\label{Eq:BScond}
\ee
The second term in the r.~h.~s. of Eq.~(\ref{Eq:BScond}) vanishes in the limit 
of $q^2\rightarrow -\infty$. It clearly indicates that negative infinite 
$q^2$ is responsible for the bound-state formation. 
Therefore, in this limit, the relative momentum squared $p^2$ 
approaches $-\gamma^2$, which must be non-zero. 
Meanwhile, the negative infinite $q^2$ turns 
out to be the infinite volume limit. 
This representation shows that although the pole condition is fulfilled in the infinite volume, 
its modification in finite volume is described by correction terms, which are 
exponentially suppressed by the size of spatial extent $L\propto q$.

Although it was pointed out how the bound-state pole
condition could be implemented in his phase-shift formula in the original 
paper~\cite{Luscher:1990ux}, another type of large $L$ expansion formula
around $q^2=-\infty$ has been explicitly derived in Ref.~\cite{Beane:2003da}. 
%
%
\begin{equation}
\Delta E_{q^2=-\infty}=-\frac{\gamma^2}{2\mu}\left[
1+\frac{12}{\gamma L}\frac{1}{1-2\gamma(p\cot \delta_{0})^{\prime}}
e^{-\gamma L}+{\cal O}(e^{-\sqrt{2}\gamma L})+{\cal O}(\gamma^2/\mu^2)
\right],
\label{Eq.Bound}
\end{equation}
where $(p \cot \delta_0)^{\prime}=\frac{d}{dp^2}(p \cot \delta_0)|_{p^2=-\gamma^2}$.
An $L$-independent term $-\frac{\gamma^2}{2\mu}$ corresponds to the binding energy 
in the infinite volume limit. 
We can learn from Eq.~(\ref{Eq.Bound}) that ``loosely bound states" 
are supposed to receive larger finite volume corrections
than those of ``tightly bound states" since the expansion parameter 
is scaled by the binding momentum $\gamma$.
Furthermore, it can be expected that the bound state of two or more particles 
has a kinematical nature similar to a single particle 
if the spatial size $L$ is much larger than the size of its compositeness, which
may be characterized by the inverse of the binding momentum.


\subsection{
Novel view from Levinson's theorem
}
\label{Sec:2-C}

\begin{figure}[b]
\begin{center}
\includegraphics[scale=0.45]{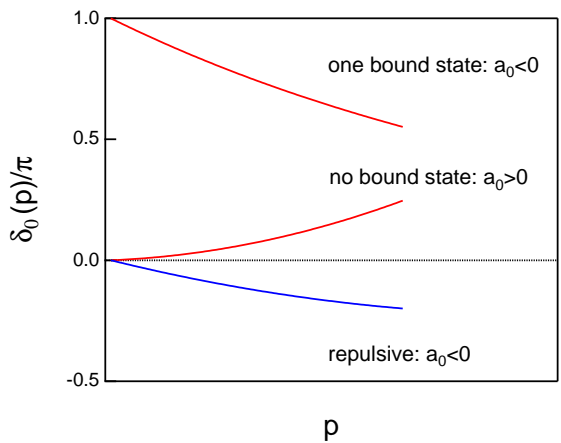}
\end{center}
\caption{A schematic figure for the scattering phase shift as a function of the relative momentum of two-particle.
}
\label{FIG:PhaseSft}
\end{figure}

At last, a crucial question arises: once the 
$S$-wave bound states are formed, what is the fate of the lowest $S$-wave 
scattering state? The answer to this question might provide 
a hint to resolve our main issue of how to distinguish between ``loosely bound states" 
and scattering states. A naive expectation from Levinson's theorem in quantum 
mechanics is that the energy shift relative to a threshold turns out to be opposite in 
comparison to the case where there is no bound state. 
Levinson's theorem relates the elastic scattering phase shift $\delta_l$ for 
the $l$-th partial wave at zero relative momentum to the total number of bound states ($N_{l}$) in a beautiful relation~\footnote{
Strictly speaking, this form is only valid unless zero-energy resonances exist.}:
%
%
\be
\delta_{l}(0)=N_{l} \pi.
\ee
Therefore, if an $S$-wave bound state is formed in a given channel, the 
$S$-wave scattering phase shift should always be positive at low energies. 
This positiveness of the scattering phase shift is consistent with a 
consequence of  the attractive interaction. 
Conversely, the $S$-wave scattering length may become negative ($a_0<0$) 
as schematically depicted in Fig~\ref{FIG:PhaseSft}.
Consequently, according to Eq.~(\ref{Eq.ScattL0})~\footnote{
In Ref.~\cite{Luscher:1985dn}, Eq.~(\ref{Eq.ScattL0}) is derived under the assumption 
that there is no bound state. However in a subsequent paper, the author stresses that Eq.~(\ref{Eq.ScattL0}) is still valid even if the bound state is formed. 
This is a consequence of the orthogonality of bound states and scattering states.}, 
possible negativeness of the scattering length
gives rise to a positive energy-shift of the lowest scattering state relative to 
the threshold energy. In other words, the lowest ($n=0$) scattering state
is pulled up into the region {\it above threshold}. 
Therefore, the spectra of the scattering states quite
{\it resembles the one in the case of the repulsive interaction}. 
If it were true, we can observe a significant difference 
in spectra above the threshold between the two systems: one has at least one 
bound state (bound system) and the other has no bound state (unbound system).

%
%
%

\section{Setup of numerical simulations}
\label{Sec:3}

\subsection{Compact Scalar QED}
\label{Sec:3-A}

To explore signatures of bound-state formation on the lattice,
we consider a bound state (positronium) between an electron
and a positron in the compact QED with scalar matter: 
%
%
\begin{equation}
S_{\rm SQED}[U,\Phi,\Psi]=S_{\rm AH}[U, \Phi]+\sum_{\rm sites}\overline{\Psi}_x D_{\rm W}[U]_{x,y} \Psi_y ,
\label{Eq:SQED}
\end{equation}
which is the compact $U(1)$ gauge theory coupled to both scalar matter (Higgs) fields
$\Phi$ and fermion (electron) fields $\Psi$. The action of ``$U(1)$ gauge + Higgs" part
is described by the compact $U(1)$-Higgs model:
%
%
\begin{equation}
S_{\rm AH}[U, \Phi]
=\beta \sum_{\rm plaq.}\left[1- \Re \{ U_{x, \mu \nu}\}\right]
-h\sum_{\rm link} \Re \{\Phi^{\ast}_x U_{x, \mu}
\Phi_{x+{\mu}}
\},
\label{Eq:AH}
\end{equation}
where $\beta=1/e^2$ and 
the constraint $|\Phi_{x}|=1$ is imposed.  In tree level,
the vacuum expectation value of the Higgs field and the photon mass are 
interpreted as $\langle \phi_{\rm higgs} \rangle \sim a^{-1}\sqrt{h}$
and $M_{\rm ph}\sim a^{-1}\sqrt{h/\beta}$ respectively~\cite{Fradkin:1978dv}.
In the Higgs phase, the Coulomb potential should be screened by the massive
photon fields:
%
%
\be
V(r)\simeq\frac{e^2}{4\pi}\frac{e^{-M_{\rm ph}r}}{r}.
\ee
%

%
%
\begin{figure}[b]
\begin{center}
\includegraphics[scale=0.4]{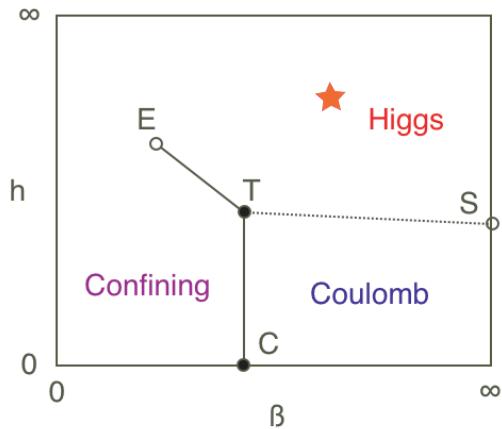}
\end{center}
\caption{Schematic
phase diagram of the compact $U(1)$-Higgs model in the fixed modulus case.
A star mark represents our simulation point as $(\beta, h)=(2.0,\;0.6)$.
}
\label{FIG:PhaseDiagram}
\end{figure}
The phase structure of the $U(1)$-Higgs model has been well studied
on the lattice. Fig.\ref{FIG:PhaseDiagram} shows a schematic
phase diagram of the compact $U(1)$-Higgs model.
There are three phases: the confinement phase, the Coulomb phase
and the Higgs phase.
The open symbols and filled symbols
represent the second-order phase transition points ({\bf E}: the end point
$\{\beta,h\}=\{0.8485(8),0.5260(9)\}$~\cite{Alonso:1992zg} 
and {\bf S}: the 4-dim XY model phase transition)
and the first-order phase transition points ({\bf T}: the triple point 
$\{ \beta, h\}\sim\{1, 0.36\}$
and {\bf C}: the pure compact $U(1)$ phase transition $\beta_c\simeq1.01$)
respectively.
Lines ET and TC represent the first order line. A dotted line TS corresponds
to the Coulomb-Higgs transition, of which the order is somewhat controversial
in the literature because of large finite size effects. 

\subsection{Monte Carlo simulation}
\label{Sec:3-B}

In this numerical study, we treat the fermion fields  in the quenched approximation. 
Therefore, for update of gauge links and Higgs fields, we simply adopt the Metropolis algorithm. 
First, the acceptance is adjusted to about 30\%. Then we use 16 hits at each link
and Higgs field update. 

Our purpose is to study the $S$-wave bound state and
scattering states through L\"uscher's finite size method,
which is only applied to the short-ranged interaction case.
Thus, we fix $\beta=2.0$ and $h=0.6$ in Eq.(\ref{Eq:AH}) to simulate 
the Higgs phase of $U(1)$ gauge dynamics, where massive photons 
give rise to the short-ranged interparticle force between an electron and a positron.
We generate $U(1)$ gauge configurations with a parameter set,
$(\beta, h)=(2.0, 0.6)$, 
on $L^3\times 32$ lattices
with several spatial sizes, $L=12,16, 20, 24, 28$ and 32.
Statistics for each volume calculation are summarized in Table~\ref{ConfVol}.

%
%
\begin{table}[htdp]
\caption{
Simulation statistics}
\begin{ruledtabular}
\begin{tabular}{c|cccccc}
Spatial size ($L$) & 12 & 16 & 20 & 24 & 28 & 32\\
\hline
\# of conf. & 960 & 1920 & 1280 & 720 & 720 & 480
\end{tabular}
\end{ruledtabular}
\label{ConfVol}
\end{table}

Once the parameters of  the compact $U(1)$-Higgs action, $(\beta, h)$, 
are fixed, the strength of an interparticle force between electrons 
should be frozen on given gauge configurations. However, if we 
consider the fictitious $Q$-charged electron, the interparticle force can be
controlled by this charge $Q$ since the interparticle force is 
proportional to (charge $Q)^2$. Within the quenched approximation, this trick 
of the $Q$-charged electron is
easily implemented by replacing  $U(1)$ link fields as
%
%
\begin{equation}
U_{x,\mu} \longrightarrow U^{Q}_{x,\mu}=\Pi_{i=1}^{Q} U_{x, \mu}
\end{equation}
into the Wilson-Dirac matrix:
%
%
\be
D_{\rm W}[U^Q]_{x,y}=\delta_{x,y}-\kappa\sum_{\mu}\left[
(1-\gamma_{\mu})U^Q_{x,\mu}\delta_{x+\mu,y}+(1+\gamma_{\mu})
U_{x-\mu,\mu}^{Q\dagger}\delta_{x-\mu,y}
\right],
\ee
where $\kappa$ is the hopping parameter.

For the matrix inversion, we use the BiCGStab algorithm~\cite{Frommer:1994vn} 
and adopt the convergence condition $|r|<10^{-15}$ for the residues. 
We calculate the electron propagators
$\langle0| \Psi(x)\overline{\Psi}(y)|0\rangle=D^{-1}_{\rm W}[U^Q]_{x,y}$
with both periodic and anti-periodic boundary conditions in the temporal
direction. Then, we adopt the averaged propagator over the boundary conditions.
This procedure provides an electron propagator with $2T$-periodicity~\cite{{Sasaki:2001nf},{Sasaki:2005ug}}.

\subsection{Spectrum of single electron} 
\label{Sec:3-C}

To evaluate a threshold energy of the electron-positron ($e^- e^+$) system, it is necessary
to calculate the electron mass nonperturbatively by the following two-point correlator,

%
%
\be
G_{e}(t; {\bf p}_n)=\frac{1}{L^6}\sum_{{\bf x},{\bf y}}{\rm Tr}\{
{\cal P}_{+} 
\langle0| \Psi({\bf x}, t)\overline{\Psi}({\bf y}, 0)|0\rangle e^{i{\bf p}_n\cdot({\bf x}-{\bf y})}\},
\label{Eq:EleCorr}
\ee
where ${\cal P}_{+}=\frac{1+\gamma_4}{2}$ and 
${\bf p}_n=\frac{2\pi}{L}{\bf n}$ with ${\bf n}\in Z^3$ for the periodic boundary
condition in spatial directions. Here, we have set the lattice spacing to unity ($a=1$).
This electron two-point correlator is {\it gauge-variant}, so
gauge fixing is required. We fix to the Landau gauge. 
However, it is well known that the pure compact $U(1)$ gauge 
theory in the Coulomb phase leads to a serious problem of the Gribov ambiguity 
in the gauge-fixing procedure. We adopt the modified iterative Landau gauge fixing, which is proposed in Ref.~\cite{Durr:2002jc}, to avoid the Gribov copy effect on 
{\it gauge-variant} electron correlators
as much as possible. Here, we remark that the Gribov ambiguity is not observed to be severe 
in the Higgs phase of compact scalar QED, where our simulations 
are performed, as is also true in the confined phase~\cite{Durr:2002jc}.

%
%
\begin{table}[htdp]
\caption{Two parameter sets ($Q$, $\kappa$) for electron fields and 
resulting rest masses of a single electron in lattice units.}
\begin{ruledtabular}
\begin{tabular}{ccl}
charge $Q$ & $\kappa$ & $M_{e}^{L\rightarrow \infty}$ \\
\hline
3 &  0.1639 &  0.479036(75)\\
4 &  0.2222 &  0.50396(59)
\end{tabular}
\end{ruledtabular}
\label{Table:PamaSim}
\end{table}
%

\subsubsection{Volume dependence of electron rest mass}
\label{Sec:3-C1}

\begin{figure}[htb]
\begin{center}
\includegraphics[angle=-90,scale=0.3]{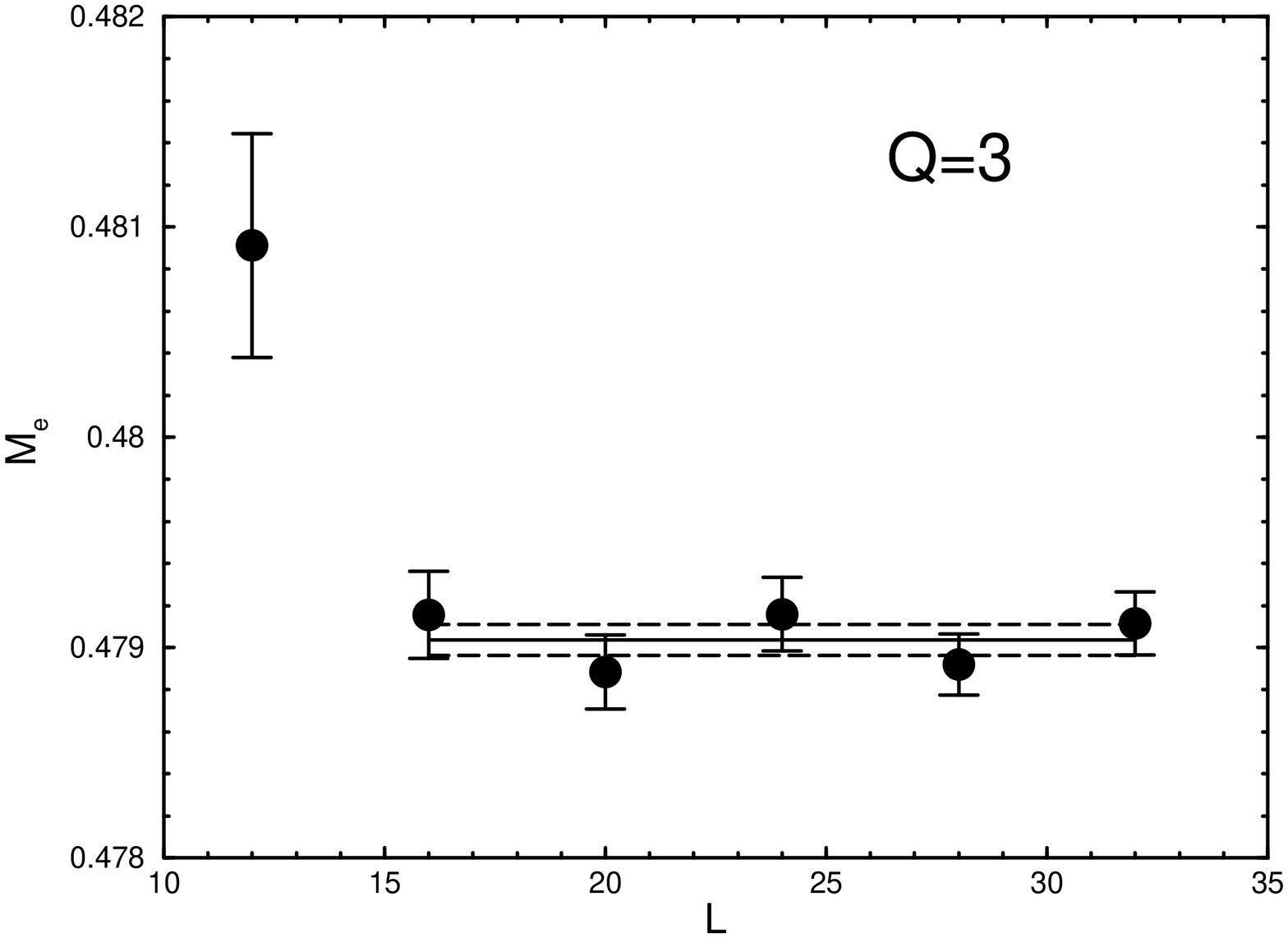}
\includegraphics[angle=-90,scale=0.3]{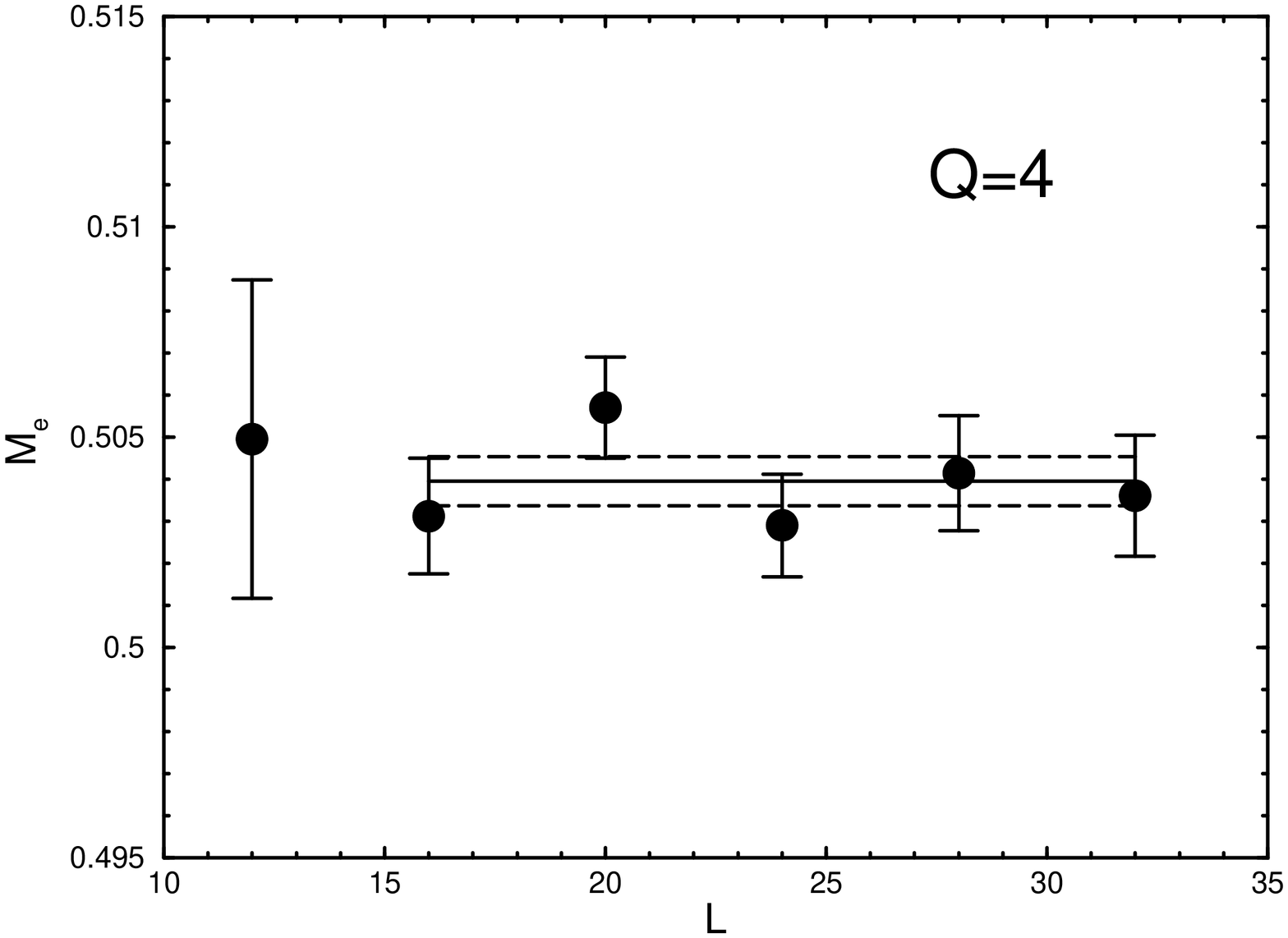}
\end{center}
\caption{Measured electron masses as a function of 
spatial size $L$ in lattice units. The left (right) panel is for three-charged (four-charged)
electron at $\kappa=0.1639$ (0.2222).
The horizontal solid line in each panel represent the value of $M_e$ 
in the infinite volume limit, which is obtained from the weighted average of five data,
with its 1$\sigma$ deviation (dashed lines). }
\label{FIG:Vol_ele}
\end{figure}

According to our previous pilot study~\cite{Sasaki:2005pc}, 
numerical simulations are performed with two parameter sets for fermion (electron) fields, $(Q, \kappa)$=(3, 0.1639) and (4, 0.2222), which are adjusted to yield almost the same 
electron masses $M_{e}\approx 0.5$ for both charges.
First, we calculate the electron mass at rest (${\bf p}_{0}=(0,0,0)$).
The electron mass is obtained by a single exponential fit, which takes 
into account the $2T$-periodicity in our simulations, 
to the two-point correlator of a single electron (\ref{Eq:EleCorr}).
In Fig.\ref{FIG:Vol_ele}, we show the volume dependence of the electron mass
for three-charged (the left panel) and four-charged (the right panel) electron fields. 
In both cases of $Q=3$ 
and 4, there is no appreciable finite size effect on the electron mass 
if the spatial lattice size $L$ is larger than 16. We take a weighted average 
of the five masses in the range $16\le L \le 32$ to evaluate values in the infinite volume limit, 
which are hereafter used in estimating a threshold energy of two-electron states.
In Fig.\ref{FIG:Vol_ele}, solid horizontal lines represent the average values taken as the infinite volume limit, together with their one standard deviation (dashed lines).
A summary of the infinite-volume values is given in Table~\ref{Table:PamaSim}.
%
%
%

%
%
\begin{table}[htdp]
\caption{Fitted masses of single electrons ($Q=3,4$) 
with zero momentum and non-zero lowest momentum
($p_1=2\pi/L$) at six different lattice volumes $L^3\times 32$.
}
\begin{ruledtabular}
\begin{tabular}{c|c@{\hspace{5mm}}c@{\hspace{5mm}}|
c@{\hspace{5mm}}c@{\hspace{5mm}}}
Spatial size& $Q=3$ & & $Q=4$ & \\
$L$ & $M_e$ & $E_e(p_1)$ & $M_e$ & $E_e(p_1)$ \\\hline
12 & 0.48091(53) & 0.67364(85) & 0.5050(38) & 0.6941(55) \\
16 & 0.47916(21) & 0.60248(34) & 0.5031(14) & 0.6235(18) \\
20 & 0.47889(18) & 0.56255(26) & 0.5057(12) & 0.5861(13) \\
24 & 0.47916(18) & 0.53953(26) & 0.5029(12) & 0.5610(13) \\
28 & 0.47892(15) & 0.52485(24) & 0.5041(14) & 0.5483(14) \\
32 & 0.47912(15) & 0.51506(27) & 0.5036(14) & 0.5388(16) \\\hline
$\infty$ & 0.479036(75) & --- & 0.50396(59) & --- \\
\end{tabular}
\end{ruledtabular}
\label{Table:ElectronEgy}
\end{table}
%

\subsubsection{Dispersion relation}
\label{Sec:3-C2}

\begin{figure}[htb]
\begin{center}
\includegraphics[angle=-90,scale=0.3]{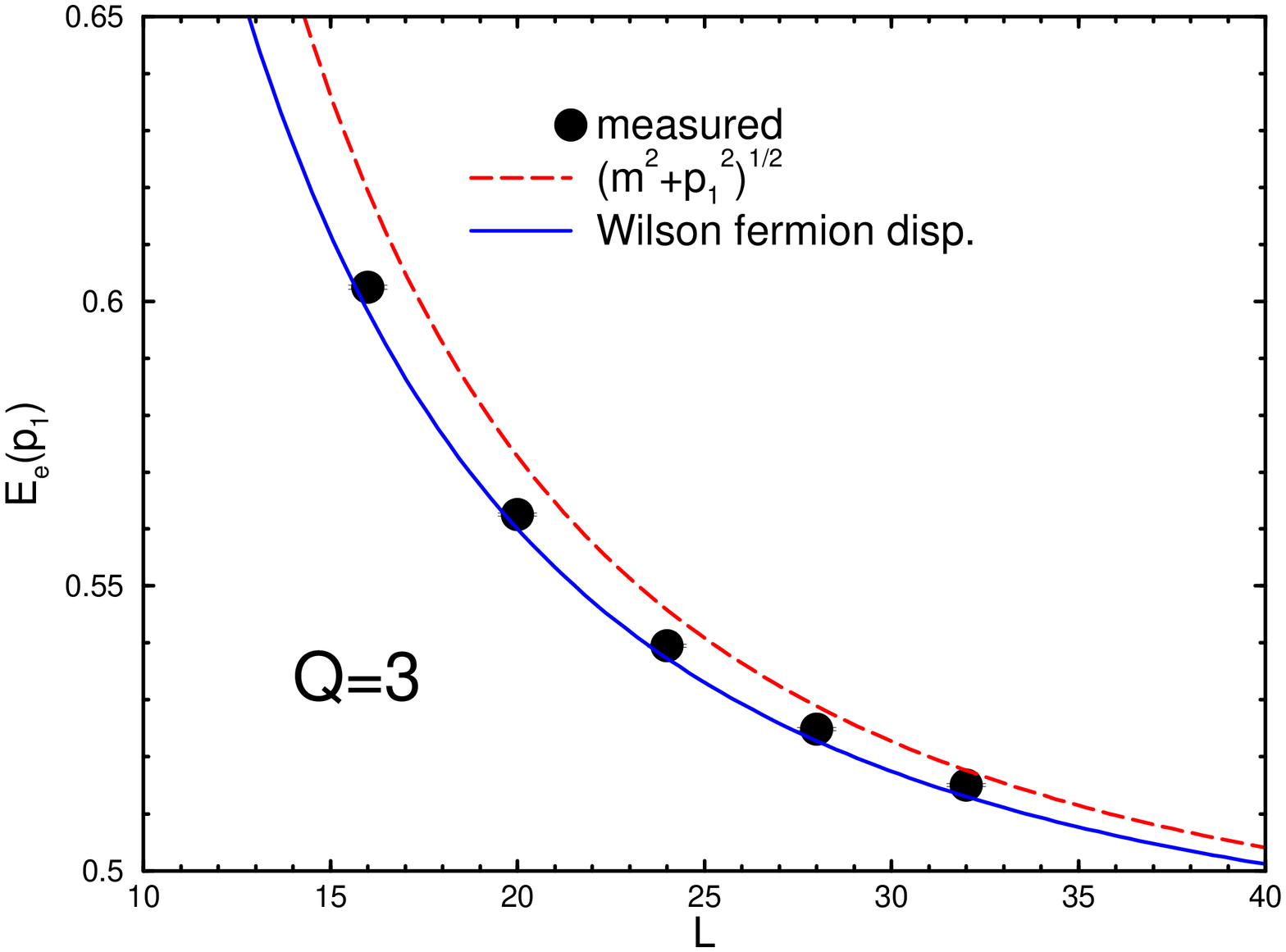}
\includegraphics[angle=-90,scale=0.3]{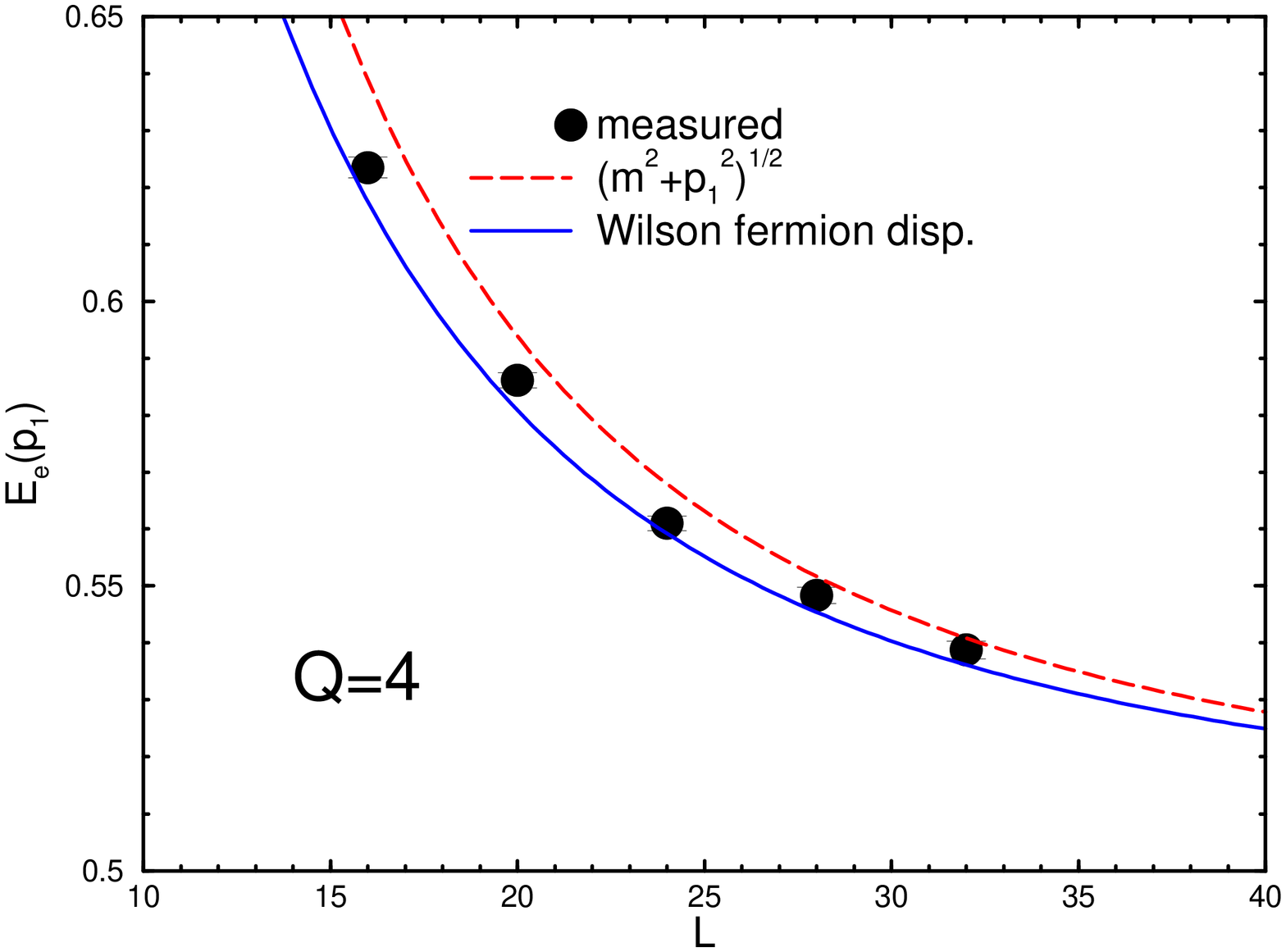}
\end{center}
\caption{
The electron energy at the non-zero lowest momentum ($p_1=2\pi/L$) as 
a function of the spatial size $L$ in lattice units. The left (right) panel for the $Q=3$ ($Q=4$) case.
Full circles represent the measured values. The solid (dashed) curves
are theoretical curves evaluated from the lattice (continuum-type) dispersion relation
with the measured rest mass.
}
\label{FIG:DisRel}
\end{figure}

Next, we examine the dispersion relation of the single electron  in our simulations
in order to study the effects of the finite lattice spacing.
We calculate the electron correlation (\ref{Eq:EleCorr}) with non-zero
lowest momentum,${\bf p}_1=\frac{2\pi}{L}(1,0,0)$, to measure
the energy level of the non-zero momentum single electron.
All measured values are tabulated in Table~\ref{Table:ElectronEgy}.
In Fig.\ref{FIG:DisRel}, we compare our measured energies $E_e(p_1)$ at 
several spatial lattices with a couple of theoretical curves, which are evaluated 
from two types of the dispersion relation with
the measured rest mass: the continuum-type dispersion relation 
%
%
\bea
E^{\rm con}_e(p_n)&=&\sqrt{M_e^2 + {\bf p}_n^2} 
\label{Eq:ConDsp}
\eea
and the lattice dispersion relation for free Wilson fermions~\cite{Carpenter:1984dd}
%
%
\bea
E^{\rm latt}_e(p_n)&=&\cosh^{-1}\left(1+
\frac{(1-\sqrt{1-\hat{\bf p}_n^2} + W)^2+\hat{\bf p}_n^2}{2(2-\sqrt{1-\hat{\bf p}_n^2} +W)}
\right),
\label{Eq:LattDsp}
\eea
where $W=e^{M_e}-1$, ${\bf p}_{n}=\frac{2\pi}{L}(n_x,n_y,n_z)$,
and $\hat {\bf p}_n^2=\sum_{k}\sin^2[\frac{2\pi}{L} n_k]$. 
The solid curves obtained from the lattice dispersion relation are clearly closer to 
the measured energies in both $Q=3$ and $Q=4$ cases.
The finite lattice spacing effects on the single electron spectra are 
not negligible even at the lowest momentum. 
Recall that the relative momentum of two particles is a key ingredient 
when we determine the scattering phase shift from Eq.(\ref{Eq.LucsherFormula}).
In this sense, the lattice dispersion relation is preferable so as to reduce the systematic 
error stemming from the lattice spacing artifact in determination of 
the relative momentum of two-particle states.
Through out this paper, we use the lattice
dispersion relation~(\ref{Eq:LattDsp}) in the analysis of the scattering phase shift through L\"uscher's formula (\ref{Eq.LucsherFormula}).

%
%
%

\subsection{Diagonalization method}
\label{Sec:3-D}

We are especially interested in the $^1S_0$ and $^3S_1$ states of the $e^{-}e^{+}$ system,
where the electron-positron bound state (positronium) could be formed even in the Higgs phase.
$^1S_0$ and $^3S_1$ positronium are described by
the bilinear pseudo-scalar operator $\overline{\Psi}_x\gamma_5 \Psi_x$ and vector 
operator $\overline{\Psi}_x\gamma_\mu \Psi_x$ respectively. 
Therefore, we may construct the four-point functions of electron-positron states 
based on the above operators.
We are interested in not only the lowest level of two-particle spectra,
but also the 2nd and 3rd lowest levels. 
In order to extract a few low-lying energy levels of two-particle system, 
we utilize the diagonalization method proposed by L\"uscher and 
Wolff~\cite{Luscher:1990ck}.
We consider three types of operators for this purpose:
%
%
\bea
\Omega_P(t)&=&\frac{1}{L^3}\sum_{\bf x}{\overline \Psi}({\bf x}, t)\Gamma \Psi({\bf x},t), \\
\Omega_W(t)&=&\frac{1}{L^6}\sum_{{\bf x},{\bf y}}{\overline \Psi}({\bf y}, t)\Gamma \Psi({\bf x},t), \\
\Omega_M(t)&=&\frac{1}{L^6}\sum_{{\bf x},{\bf y}}{\overline \Psi}({\bf y}, t)\Gamma \Psi({\bf x},t)e^{i{\bf p}_1\cdot({\bf x}-{\bf y}) },
\label{Eq:ThreeOps}
\eea
where ${\bf p}_1=\frac{2\pi}{L}(1,0,0)$ and 
$\Gamma=\gamma_5$  ($\gamma_{\mu}$) for the $^1S_0$  ($^3S_1$) $e^- e^+$ state. 
The first operator corresponds to a simple local-type 
operator where only the total momentum of two particles is fixed to be 
zero, but both the electron and the positron can carry non-zero 
relative momentum under the total momentum conservation.
The second operator projects both the electron and the positron
onto zero momentum, while the relative momentum of the $e^- e^+$ system
is constrained to the non-zero lowest momentum ($p_1=|{\bf p}_1|=\frac{2\pi}{L}$) 
in the third operator.
Therefore, we can expect that each type of operators has better 
overlap to a specific two-particle state: $n=0$ and $n=1$ scattering states 
have strong  overlap with $\Omega_W$ and $\Omega_M$ respectively, while
the bound state has the better overlap with $\Omega_P$ than 
$\Omega_W$ and $\Omega_M$.

We construct the $3\times 3$ matrix correlator from above three operators
%
%
\be
G_{ij}(t)=\langle 0|\Omega_{i}(t)\Omega^{\dagger}_j(0)|0\rangle
\label{Eq:MatCorr}
\ee
and then employ a diagonalization of a transfer matrix $M(t, t_0)$, which
is defined by
%
%
\be
M(t, t_0)=G(t_0)^{-1/2}G(t)G(t_0)^{-1/2},
\label{Eq:TransMat}
\ee
where $t_0$ is a reference time-slice. If only three states are propagating 
in the region $t>t_0$, the energies of three two-particle states $E_{\alpha}$
($E_2>E_1>E_0$) are given by the eigenvalues of $M(t,t_0)$:
%
%
\be
\lambda_{\alpha}(t,t_0)=e^{-(t-t_0)E_{\alpha}}\;\;\;(\alpha=0,1,2),
\ee
where $E_{\alpha}$ is independent of $t_0$. 
An assumption that three low-lying states become effectively dominant 
for an appropriately large time-slice $t_0$, can be determined by checking the sensitivity
of $E_{\alpha}$ with respect to variation of the reference time-slice $t_0$.

In this study, the random noise method is employed to calculate 
$\Omega_p$ source operators in Eq.~(\ref{Eq:MatCorr})
with the number of noises taken to be one. Technical details of this method 
are described in Ref.~\cite{{Aoki:2002ny},{Yamazaki:2004qb}}.
We note that all contributions from disconnected diagrams 
in Eq.~(\ref{Eq:MatCorr}) are simply ignored
in our numerical calculations.

\section{Numerical results in the  $^{1}S_0$ channel}
\label{Sec:4}

In this section, we focus on numerical results in the $^{1}S_0$ channel 
of the $e^{-}e^{+}$ system. Results obtained in the $^{3}S_1$ channel will 
be separately discussed in the next section.

\subsection{Ground state of $^{1}S_0$}
\label{Sec:4-A}

\begin{figure}[htb]
\begin{center}
\includegraphics[angle=-90,scale=0.3]{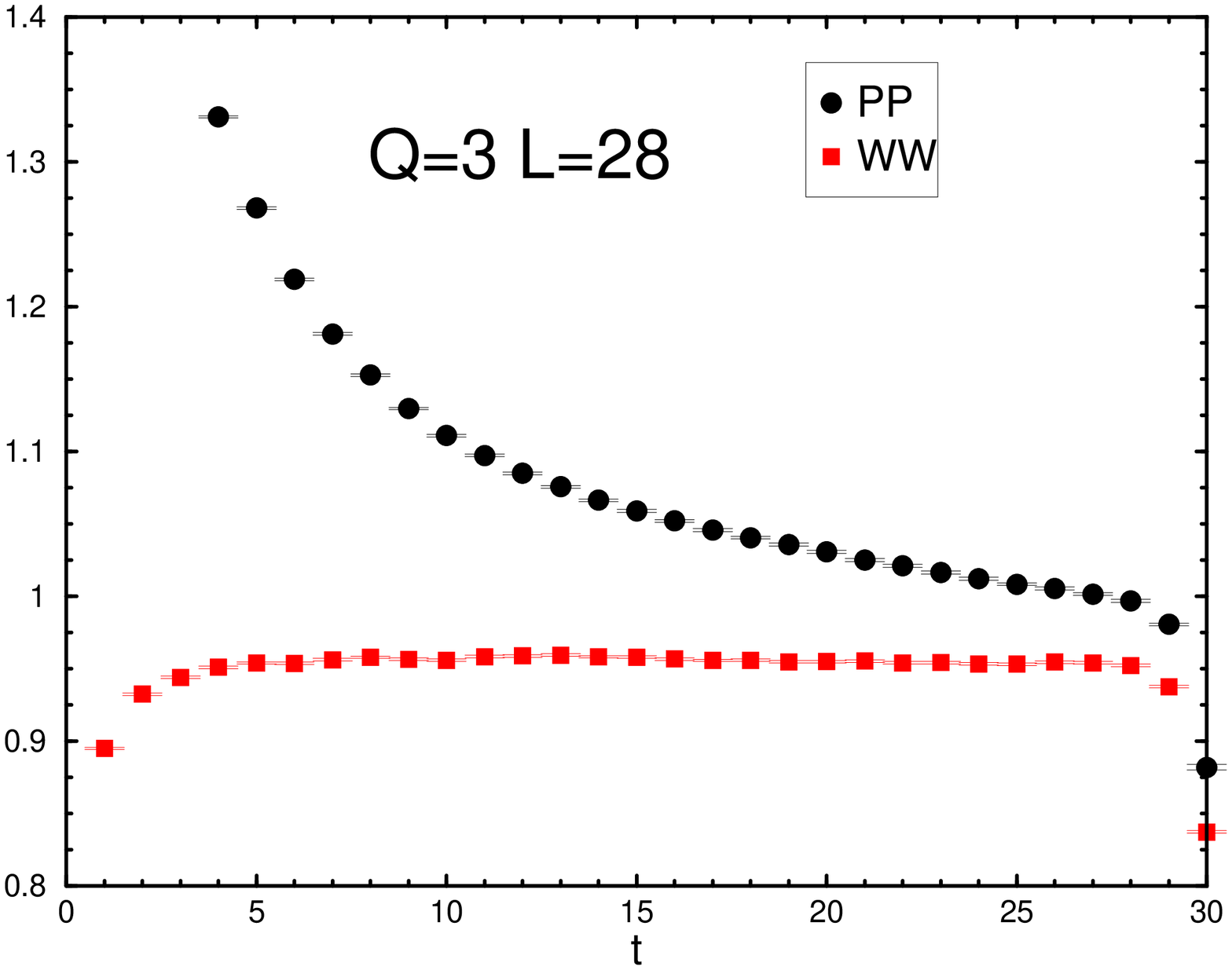}
\includegraphics[angle=-90,scale=0.3]{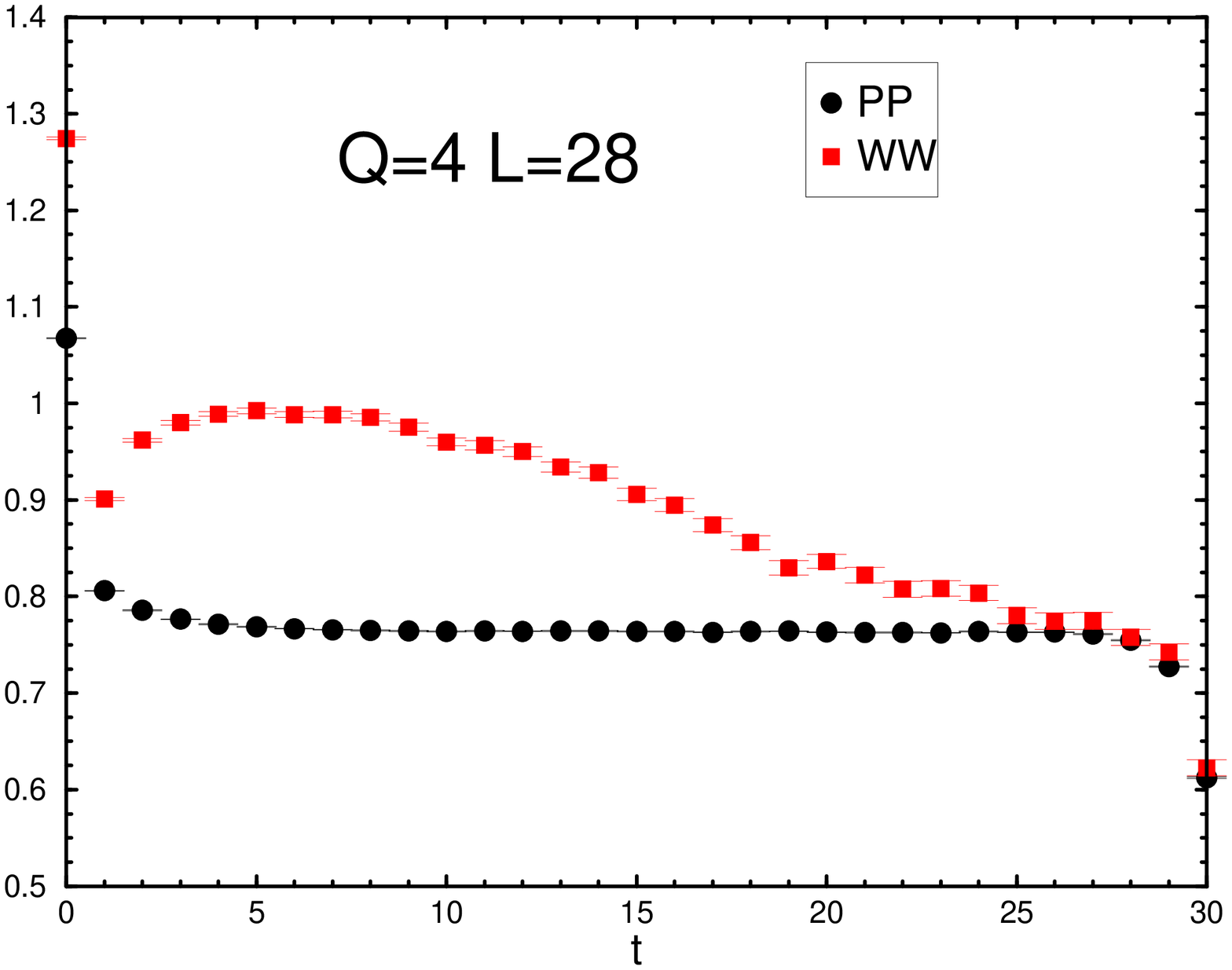}
\end{center}
\caption{
The effective masses in the $^{1}S_{0}$ channel on the lattice with
$L=28$ as a function of the time-slice $t$ in lattice units.
The left (right) panel is for $Q=3$ ($Q=4$) electron fields.
Full circles (full squares) symbols are obtained from the $PP$ ($WW$) correlator. }
\label{FIG:OpDep}
\end{figure}

Let us begin with the ground state in the $^{1}S_0$ channel. 
It is not necessary to employ the diagonalization method for the spectroscopy 
of the ground state. We first show the effective mass plot for two diagonal components
of the $3\times 3$ matrix correlator. Figs.~\ref{FIG:OpDep} show the effective mass
of the $PP$ correlator and the $WW$ correlator in simulations at spatial extent 
$L=28$ for $Q=3$ (left panel) and $Q=4$ (right panel). At a glance, there are apparent
operator dependencies. A very clear plateau appears for the $WW$ correlator 
in the $Q=3$ case, while the same quality shows up for the $PP$ correlator 
in the $Q=4$ case. This drastic change in operator dependence strongly
suggests a signature of bound-state formation in the $Q=4$ case, since
the $WW$ correlator is expected to have a large overlap with 
the lowest ($n=0$) scattering state rather than the bound state.
In addition, the energy of the $^1S_0$ ground state is close to the threshold 
energy ($2M_e \simeq 0.958$) in the case of $Q=3$, while there is 
a large energy gap between the ground state energy
and the threshold energy ($2M_e \simeq 1.008$) in the case of $Q=4$. 
Therefore, we may naively conclude that the ground state 
in $Q=4$ is the $^1S_0$ positronium state .

\begin{figure}[htb]
\begin{center}
\includegraphics[angle=-90,scale=0.3]{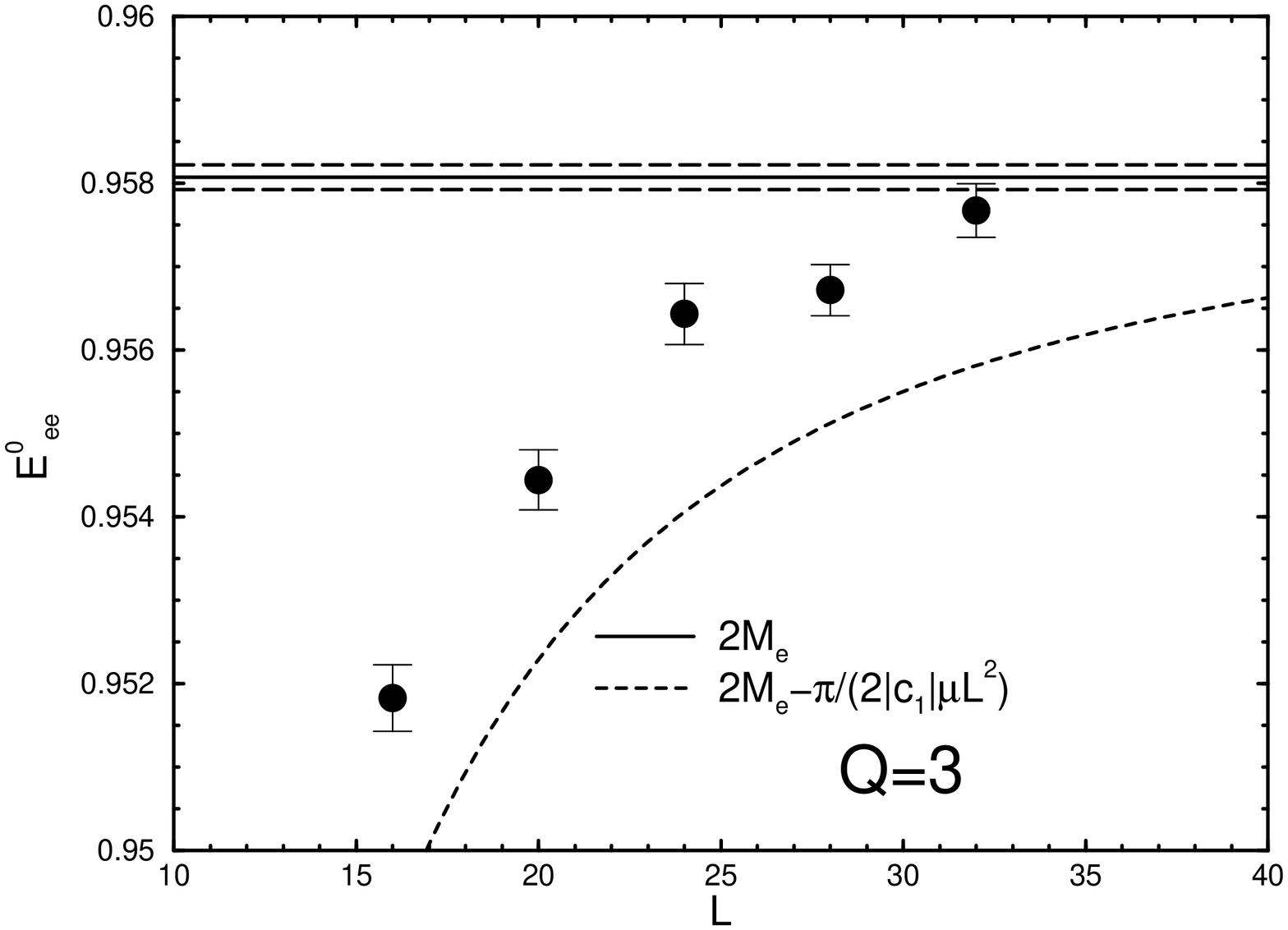}
\includegraphics[angle=-90,scale=0.3]{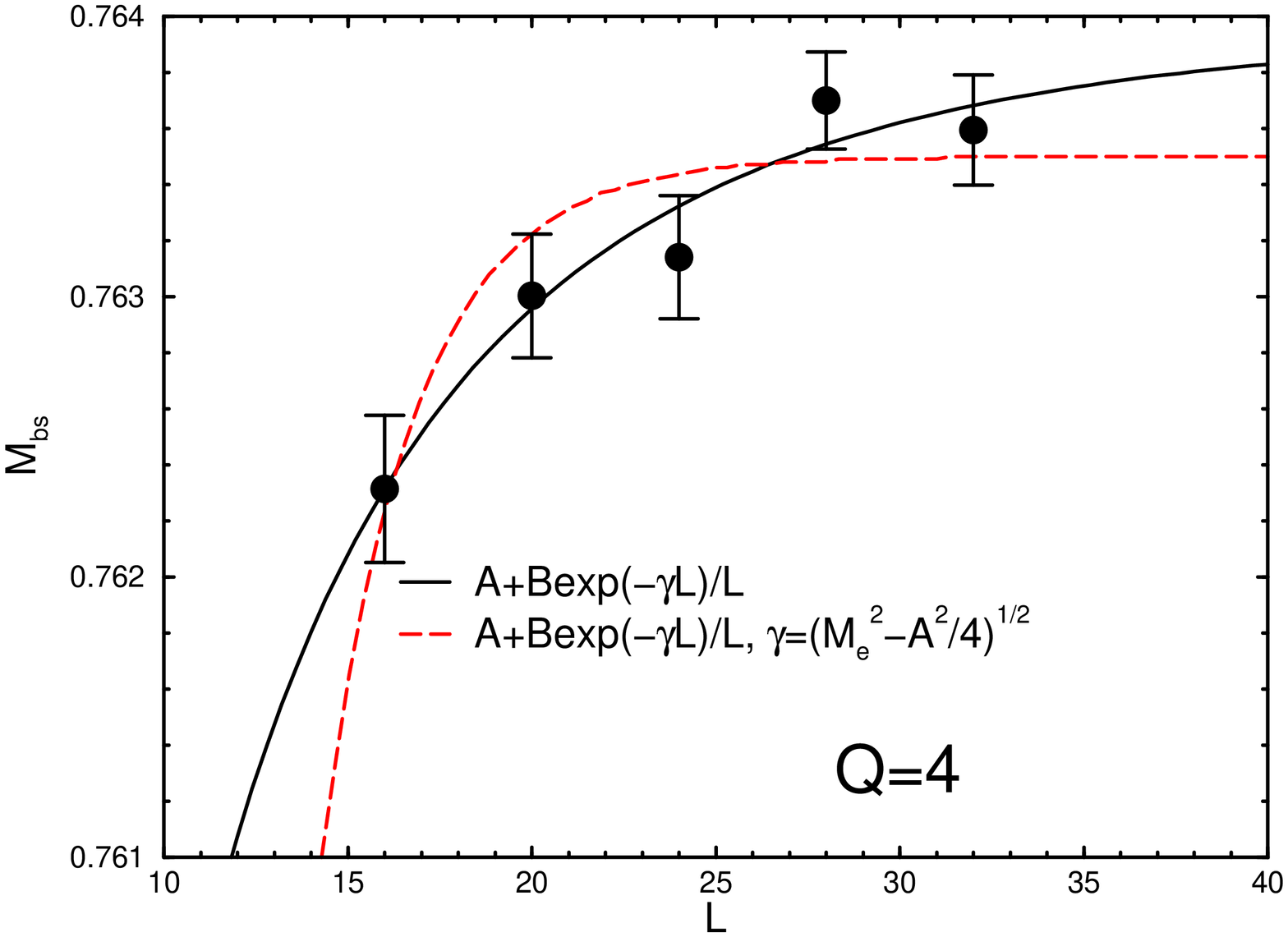}
\end{center}
\caption{Energies of ground states in the $^{1}S_{0}$ channel of the 
$e^- e^+$ system as a function of spatial lattice size $L$.
In the left panel ($Q=3$), full circles are measured energies at each lattice size.
Horizontal lines represent the threshold energy $2M_e$ and its 1 standard
deviation (dashed lines). The dashed curve shows the lower boundary for 
the convergence of the large $L$ expansion formula around $q^2=0$ 
as Eq.~(\ref{Eq.ScattL0}). 
In the right ($Q=4$) panel, full circles are measured 
energies at each lattice size, and the solid (dashed) curve is the fitting result by using
a form $E=A+B\exp(-\gamma L)/L$, which is inspired by the large $L$ expansion
formula at infinite negative $q^2$ as Eq.~(\ref{Eq.Bound}),  
with (without) a constraint between two parameters
($A$ and $\gamma$). 
}
\label{FIG:Vol_Grd}
\end{figure}

To make a firm conclusion on this point, we next show the volume dependence 
of the ground state energy in Fig.~\ref{FIG:Vol_Grd}. In the left panel ($Q=3$), 
we plot ground state energies measured at each $L$ together with 
the threshold energy as horizontal lines, which are estimated by $2M_e$
and its one standard deviation. An upward tendency of the $L$ dependence
toward the threshold energy is clearly observed as spatial size $L$ increases. 
We also include a lower bound for the asymptotic solution of the scattering state. 
All data points are located well above this lower bound. 
From those observations, we can conclude that the observed 
ground state in the $Q=3$ is definitely the lowest ($n=0$) scattering state.

On the other hand, in the right panel ($Q=4$), all data points are located far 
below the threshold energy 
and also the lower bound for the asymptotic solution of the scattering state.
Indeed, data are well fitted by the form:
\be
E(L)=A+\frac{B}{L} \exp(-\gamma L),
\label{Eq:ExpFit}
\ee
which is inspired by the asymptotic solution of the bound state, Eq.~(\ref{Eq.Bound}). 
Finding $A\ne 2M_e$ directly indicates that the energy gap from the threshold 
remains finite in the infinite volume limit.
We perform two types of fitting procedure with this form. 
First,  a full three-parameter fit is employed. Second, 
we take into account a relation between two parameters $A$ and $\gamma$
according to Eq.~(\ref{Eq.Bound}). The parameter $A$ is 
the value of ground-state energy in the infinite volume, while 
$\gamma$ corresponds to the binding momentum related to 
the pole location of the $S$-matrix as $p^2=-\gamma^2(<0)$. 
Therefore, an explicit constraint between two parameters $A$ and $\gamma$ 
can be imposed through the relation $\gamma=\sqrt{M_e^2-A^2/4}$
referred to the measured electron mass $M_e$.
Then a two-parameters fit is carried out.
All fitting results are tabulated in Table.~\ref{FitBoundState}.
Either procedure provides reasonable fits with about $\chi^2/{\rm d.o.f.}\sim 1$.
The resulting values of $A$ in both fits are approximately consistent 
with each other, while some differences show up in other parameters. 

We here stress that the obtained value of $A$  
is significantly far from the threshold value $2 M_e\simeq 1.008$ and therefore
the energy gap $\Delta E=A-2M_e$ clearly remains finite in the infinite volume limit.
A bound state of an electron and positron is certainly formed in simulations with charge-four electrons.

%
%
%

%
%
%

%
%
\begin{table}[htdp]
\caption{
Fitting results for the $^{1}S_{0}$ ground state of $Q=4$ electron fields 
using a fitting form of Eq.~(\ref{Eq:ExpFit}). 
In the table, ``Fit 1"  and ``Fit 2" stand for  
the fully three-parameters fit and the two-parameters fit with a constraint between
$A$ and $\gamma$.}
\begin{ruledtabular}
\begin{tabular}{l|cccc}
& $A$ & $B$ & $\gamma$ & $\chi^2/{\rm d.o.f.}$ \\
\hline
Fit 1&0.76395(26) & -0.78(30) & 0.068(28) & 0.87    \\
Fit 2& 0.76350(11) & -3.92(89) & constrained as $\sqrt{M_e^2-A^2/4}$   & 1.58 
\end{tabular}
\end{ruledtabular}
\label{FitBoundState}
\end{table}
%

\subsection{Excited state of $^{1}S_0$}
\label{Sec:4-B}

In the previous subsection, we confirm that the simulations with three-charge
electrons provide the purely elastic scattering system without bound states
(unbound system), while four-charge electrons give rise to at least 
one bound state as the ground state in the $^{1}S_{0}$ channel of 
the $e^{-}e^{+}$ system (bound system). 
We now can explore the difference of spectra between the unbound system
($Q=3$) and the bound system ($Q=4$).

\begin{figure}[b]
\begin{center}
\includegraphics[angle=-90,scale=0.3]{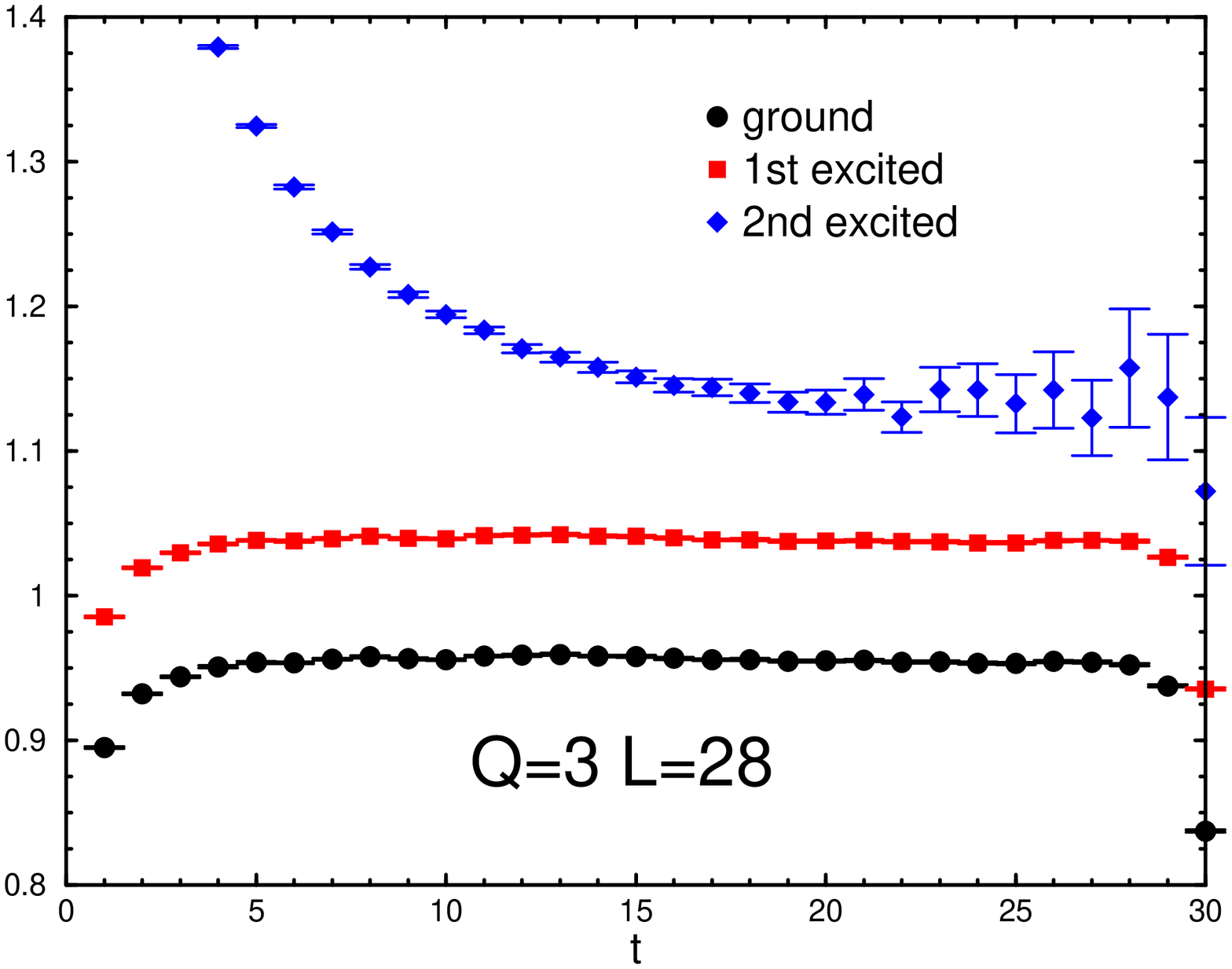}
\includegraphics[angle=-90,scale=0.3]{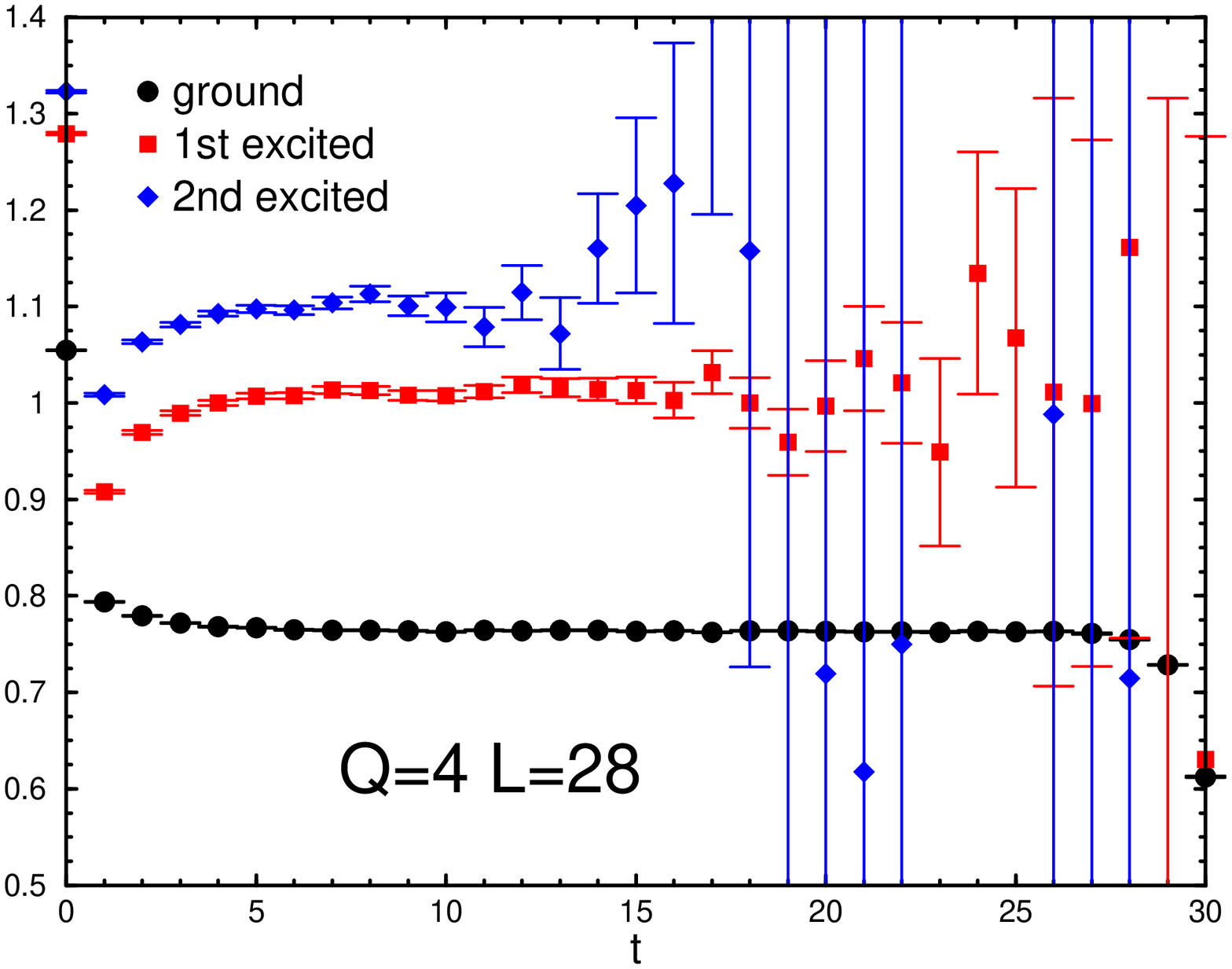}
\end{center}
\caption{The effective mass plots for each eigenvalue $\lambda_{\alpha}(t,t_0)$
of the transfer matrix defined in Eq.~(\ref{Eq:TransMat}) at a reference time-slice
$t_0=7$ on the lattice with $L=28$. Full circles, squares and diamonds represent  
the ground state ($\alpha=0$), the first excited state ($\alpha=1$) 
and the second excited state ($\alpha=2$).
The left (right) panel is for $Q=3$ ($Q=4$).}
\label{FIG:Diag}
\end{figure}

We calculate the eigenvalues $\lambda_{\alpha}(t, t_0)$ of the transfer matrix
$M(t, t_0)$ for $t_0=7$ at all $L$ except for $L=32$ where $t_0=9$ is chosen.
First we show the effective mass plots for all three eigenvalues 
$\lambda_{\alpha}(t, t_{0}=7)$ in simulations at $L=28$ in Fig.~\ref{FIG:Diag}.
The diagonalization method with our chosen three operators successfully 
separates the first excited state and the second excited state from 
the ground state. 

In the left panel ($Q=3$), the lowest and the second lowest states show 
very clear plateaus started from $t\simeq 5$, which is earlier than
our reference time-slice $t_0$. The ground state and the first excited state
correspond to the lowest ($n=0$) scattering state and the second lowest 
($n=1$) scattering state.
Those two-particle energies $E_{ee}^0$ and $E_{ee}^1$ are close to twice the single electron
energies, $2E_e(p_0)=2M_e$ and $2E_e(p_1)$, respectively. Needless to say, the energy
of the lowest state in the diagonalization method is consistent with the energy 
obtained by the $WW$ correlator. By detail analysis of 
the spectral amplitude (see, Table~\ref{Table:L28normAmpl} and Appendix B), 
we confirm that the $WW$ correlator 
and the $MM$ correlator are dominant in $\lambda_0$ and $\lambda_1$ 
respectively as expected. Although the effective mass of the third eigenvalue $\lambda_2$
gradually approaches some plateau around $t\approx 20$, statistical errors 
becomes large in the plateau region. $\lambda_2$
is dominated by the $PP$ correlator, which can overlap with any relative momentum 
scattering state, so that the contamination from higher relative momentum 
($n\ge 3$) scattering states is inevitable in the earlier time-slice.

For the bound system ($Q=4$), all three eigenvalues show 
clear plateaus started from $t\approx t_0=7$ in the effective mass plot.
Again, the energy of the lowest state in the diagonalization method agrees well 
with the one obtained from the $PP$ correlator. The obtained eigenvectors also
indicate that the $PP$ correlator is dominant in the $\lambda_0$ eigenvalue, 
while the second and third  eigenvalues are mostly composed of the $WW$ correlator
and the $MM$ correlator, respectively. As we mentioned, the $PP$ correlator possibly 
has overlap with any relative momentum scattering states. However, here,
the $PP$ correlator has dominant overlap with the bound state 
as shown in Table~\ref{Table:L28normAmpl}. This is because
the spectral weight of two-particle states relative to the single particle state, such as 
a bound state, could be suppressed in the $PP$ correlator by an inverse factor of the volume, $1/L^3$~\footnote{
This particular feature is pointed out in Refs~\cite{{Yamazaki:2002ir},{Mathur:2004jr}}.
However, a more rigorous argument regarding 
the normalization of both interacting and non-interacting 
two-particle states in finite volume can be found in Ref.~\cite{Lin:2001ek}.
We recapitulate the main point in Appendix B.}.
%

Finally, a summary table of low-lying spectra in the $^1S_0$ channel
in simulations at $L=28$ is given in Table \ref{Table:L28spect}.

%
%
\begin{table}[htdp]
\caption{Summary of the normalized spectral weights
$(A_{\alpha})_{i}$ in the $^{1}S_{0}$ channel on the lattice with $L=28$.
A definition of the normalized spectral weights is described in Appendix B.
}
\begin{ruledtabular}
\begin{tabular}{c|clll}
& & ground state & 1st excited state & 2nd excited state \\
charge $Q$ & operator $i$ & $\alpha=0$ & $\alpha=1$
& $\alpha=2$ \\
\hline
3 & $P$ & 0.02166(72)  & 0.1324(96)  & 0.846(10) \\
    & $W$ & 0.99735(37) & 0.00177(11) & 0.00088(31) \\
    & $M$ & 0.001155(48) & 0.9907(11)  & 0.0081(11) \\
\hline
4 & $P$ & 0.9597(83)  &0.0064(21) &0.0339(81) \\
   & $W$ & 0.01567(69) &0.9841(14) &0.00018(78) \\
   & $M$ &0.0521(15) & 0.0023(17) &0.9456(24) \\
\end{tabular}
\end{ruledtabular}
\label{Table:L28normAmpl}
\end{table}
%

%
%
\begin{table}[htdp]
\caption{Summary of low-lying spectra in the $^{1}S_{0}$ channel on the lattice with $L=28$.
}
\begin{ruledtabular}
\begin{tabular}{c|rlc}
charge $Q$ &eigenstate $\alpha$  
& energy $E_{\alpha}$  & kinds of state\\
\hline
3 &ground state ($\alpha=0$) 
                                                          &$E_{ee}^0$=0.95672(31) & $n=0$ scattering state\\
    &1st excited state ($\alpha=1$) 
                                                          &  $E_{ee}^1$=1.04062(41) & $n=1$ scattering state\\
\hline
4 &ground state ($\alpha=0$)      
                                                                & $M_{\rm bs}$= 0.76370(17)& bound state\\
&1st excited state ($\alpha=1$) 
                                                                &$E_{ee}^{0}$=  1.0119(26)& $n=0$ scattering state\\
&2nd excited state ($\alpha=2$) 
                                                                &$E_{ee}^{1}$=  1.1044(53)& $n=1$ scattering state
\end{tabular}
\end{ruledtabular}
\label{Table:L28spect}
\end{table}
%

%
%
%

\subsection{
Distinctive signatures of bound-state formation
}
\label{Sec:4-C}
\subsubsection{Sign of energy shift}
\label{Sec:4-C-1}
Suppose that L\"uscher's finite size method reflects all of the essential nature of the 
scattering theory in the quantum mechanics; formation of the $S$-wave
bound state is accompanied by an abrupt sign change of the scattering length.
Thus, we can expect that the second lowest energy state, which corresponds 
to the lowest ($n=0$) scattering state, should be located near and above the threshold 
energy ($2M_e$) if a bound state is formed. This is quite 
in contrast with the case if there is no bound state:
the second lowest energy state, which should be the $n=1$ scattering state, 
is located near below (above) the energy level of non-interacting two-particle system 
with non-zero lowest momentum as $2E_e(p_1)$ in the attractive (repulsive) channel. 

\begin{figure}[htb]
\begin{center}
\includegraphics[angle=-90,scale=0.3]{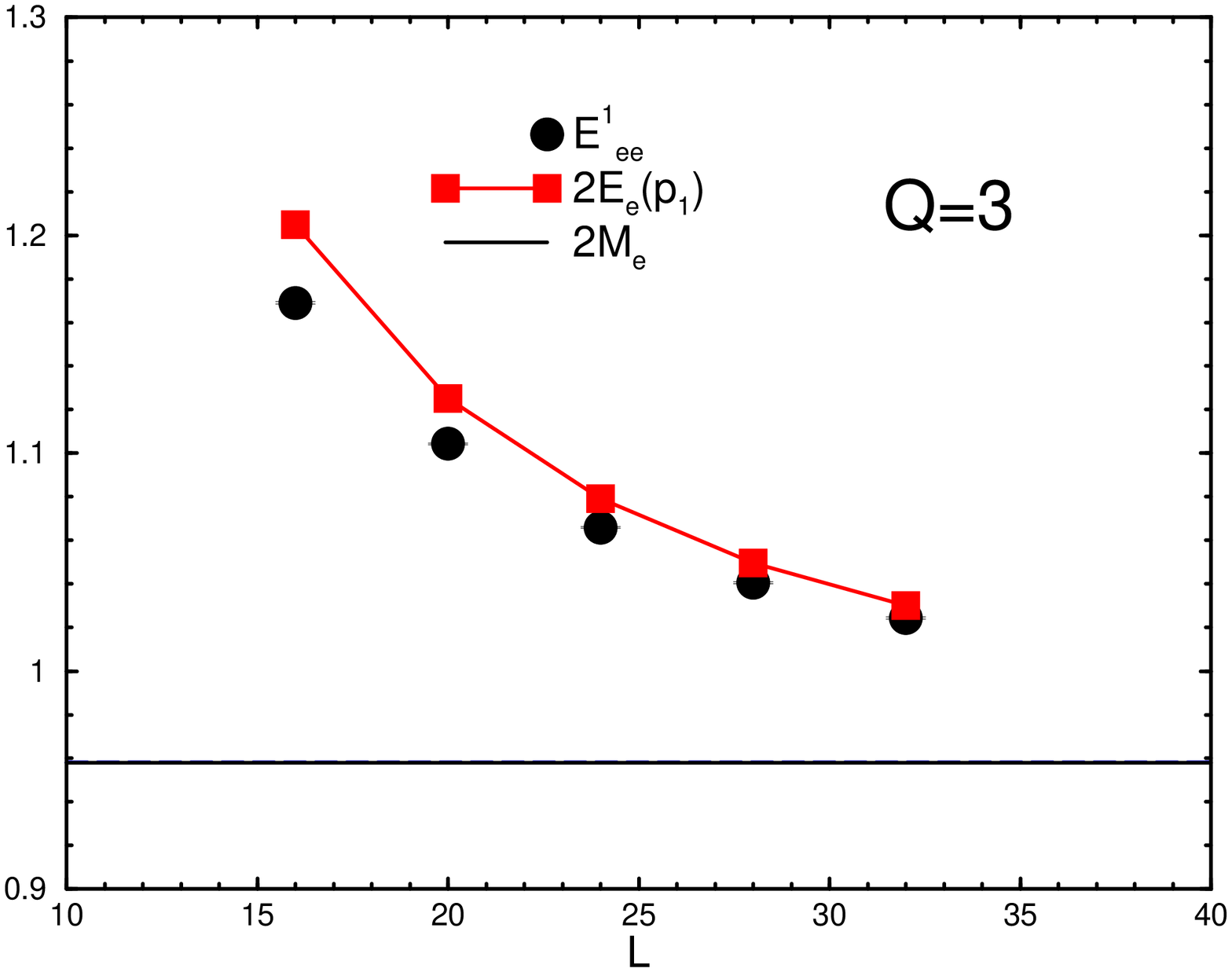}
\includegraphics[angle=-90,scale=0.3]{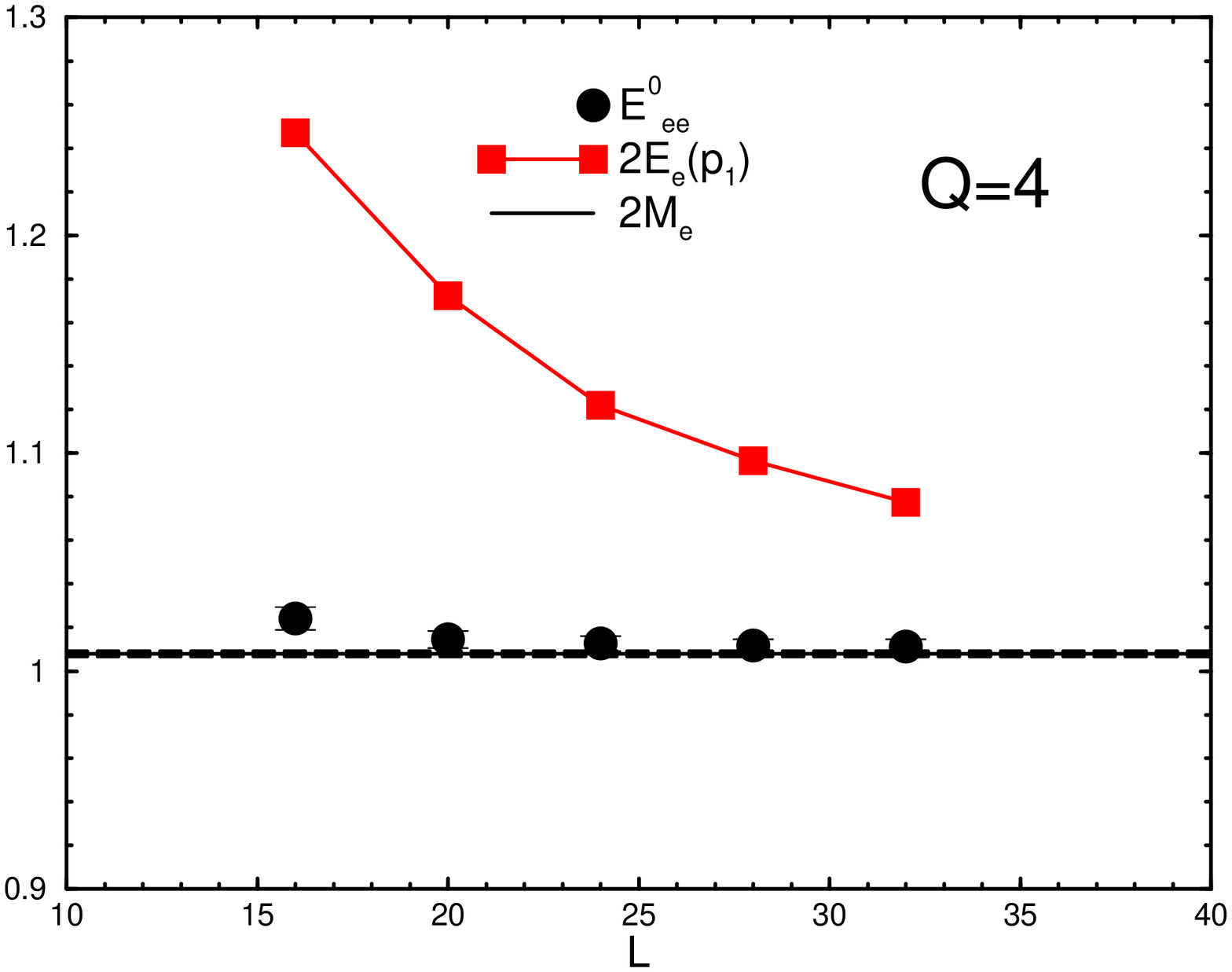}
\end{center}
\caption{Energies of the first excited state in the $^{1}S_{0}$ channel
of the $e^- e^+$ system as a function of spatial lattice size $L$. 
The left (right) panel is for $Q=3$ ($Q=4$). The horizontal line
represents the threshold energy determined by $2M_e$.
Full circles are measured energies for the first excited state.  
Solid curves with full squares shows twice of the single electron
energy with non-zero smallest momentum $p_1=2\pi/L$.
}
\label{FIG:Vol_Exc}
\end{figure}

Here we show our observed $L$-dependence of the energy level of the second lowest
state in Fig.~\ref{FIG:Vol_Exc}. The data plotted appear in Table~\ref{Table:1S0spect}. 
In the left panel ($Q=3$), measured energy levels
are very close to the $n=1$ threshold energy, which is given by twice the single 
electron energy at non-zero lowest momentum $p_1$. 
As the spatial size $L$ increases, the energy levels
approach this $n=1$ threshold energy from below.
This is consistent with a behavior of the $n=1$ scattering state
predicted by Eq.~(\ref{Eq.ScattL1}) for the weakly attractive interaction 
without bound states.  
Therefore, one can identify the second lowest energy state as the $n=1$ scattering 
state for $Q=3$.

In the right panel ($Q=4$), an expected feature comes out.
The horizontal line represents the $n=0$ threshold energy estimated by
twice the electron rest mass. Clearly, the energy levels of the second lowest
state approach this $n=0$ threshold energy {\it from above}. 
The energy shift from the threshold vanishes as the spatial size $L$ increases.
Therefore, the second lowest energy state must be the $n=0$ scattering state.
It is worth emphasizing that the sign of $\Delta E = E^{0}_{ee}-2M_e$ is opposite in 
the case of $Q=3$ where there is no bound state. Of course, this sign is directly 
related to the sign of the $S$-wave scattering length. Thus, our numerical simulations
show that formation of the $S$-wave bound state is really accompanied by
an abrupt sign change of the scattering length. 

\begin{figure}[htb]
\begin{center}
\includegraphics[angle=-90,scale=0.3]{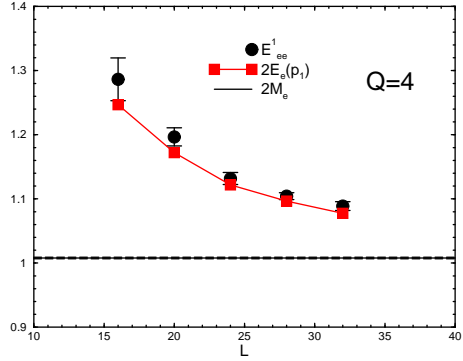}
\end{center}
\caption{
Energies of the second excited state in the $^{1}S_{0}$ channel
of the $e^- e^+$ system as a function of spatial lattice size $L$
for $Q=4$.
All symbols are defined as in Fig.~\ref{FIG:Vol_Exc}.}
\label{FIG:Vol_2ndExc}
\end{figure}

Furthermore, in Fig.~\ref{FIG:Vol_2ndExc},  
the volume dependence of the energy level of the third lowest state
in the $Q=4$ case shows the ``repulsive"  feature as the $n=1$ scattering state
even in the attractive channel. This is attributed to the consequence of Levinson's
theorem, which allows the case, $\tan \delta_0 < 0$, for the attractive interaction.

What is surprising here is that one of the most important features, namely 
Levinson's theorem, in the quantum scattering theory is inherited 
in L\"uscher's finite size formula.
Meanwhile, we realize what is a proper signature of bound state formation in finite volume
on the lattice. Even in a single simulation at fixed $L$, we can distinguish the near-threshold
bound state from the lowest ($n=0$) scattering state through determination
of whether the second lowest state appears just above the threshold 
or near the $n=1$ energy level of non-interacting two-particle states.

%
%
\begin{table}[htdp]
\caption{Energies of the ground state, the first excited state and the second excited state
(only for $Q=4$), which are obtained by a single cosh fit with $2T$ periodicity, 
in the $^{1}S_{0}$ channel of the $e^- e^+$ system
at five different lattice volumes $L^3 \times 32$.}
\begin{ruledtabular}
\begin{tabular}{c||c@{\hspace{5mm}}c@{\hspace{5mm}}|
c@{\hspace{5mm}}c@{\hspace{5mm}}c@{\hspace{5mm}}}
Spatial size& $Q=3$ & & $Q=4$ & &\\
$L$ & $E_{ee}^0$ & $E_{ee}^1$
    & $M_{\rm bs}$ & $E_{ee}^0$ & $E_{ee}^1$ \\\hline
16 & 0.95183(40) & 1.16903(55) 
   & 0.76231(26) & 1.0242(52)  & 1.287(33)\\
20 & 0.95445(36) & 1.10435(36) 
   & 0.76300(22) & 1.0145(40)  & 1.197(14)\\
24 & 0.95644(37) & 1.06590(36) 
   & 0.76314(22) & 1.0126(34)  & 1.1318(93)\\
28 & 0.95672(31) & 1.04062(41) 
   & 0.76370(17) & 1.0119(26)  & 1.1044(53)\\
32 & 0.95767(32) & 1.02440(49) 
   & 0.76359(20) & 1.0113(32)  & 1.0888(68)
\end{tabular}
\end{ruledtabular}
\label{Table:1S0spect}
\end{table}
 
%
%
%

%
%
%

\subsubsection{
Bound-state pole condition 
}
\label{Sec:4-C-2}

As we discussed in Sec.~\ref{Sec:2-B}, 
the formation condition of the $S$-wave bound state, $\cot \delta_0(p) = i$,
is definitely implemented in L\"uscher's phase-shift formula (\ref{Eq.LucsherFormula})
at {\it negative infinite} $q^2$, which corresponds to the limit of $L \rightarrow \infty$.
According to the original paper~\cite{Luscher:1990ux}, 
for {\it negative} $q^2$, 
we introduce the phase $\sigma_0(\kappa)$, which is defined
by an analytic continuation of $\delta_0$ into the complex $p$ plane 
through the relation
%
%
\be
\tan \sigma_0(\kappa)=-i\tan \delta_0(p),
\ee
where $\kappa=-ip$. Therefore, the bound-state pole condition 
in the infinite volume reads 
%
%
$\cot \sigma_0(\gamma)= -1$
%
for the binding momentum $\gamma$~\cite{Luscher:1990ux}.
As we pointed out in Sec.~\ref{Sec:2-B},
the finite volume correction on this condition
is exponentially suppressed by the spatial extent $L$
in a finite box $L^3$:
\bea
\lim_{\kappa \rightarrow \gamma}
\cot \sigma_0(\kappa) 
&=& -1 +\sum_{\nu=1}^{\infty}\frac{N_{\nu}}{\sqrt{\nu} L\gamma}e^{-\sqrt{\nu}L\gamma} \\
&=& -1 + 
\frac{6}{L\gamma}\left[e^{-L\gamma}+{\cal O}(e^{-\sqrt{2}L\gamma})\right],
\label{Eq:BScondExp}
\eea
where the factor $N_{\nu}$ is the number of integer vectors ${\bf n}\in Z^3$
with $\nu={\bf n}^2$.  
Therefore, if the bound state is formed, we may observe 
the phase $\sigma_0$ satisfies $\lim_{\kappa\rightarrow \gamma} 
\sigma_0(\kappa)=-\frac{\pi}{4}-\varepsilon_L$
where $\varepsilon_L(>0)$ vanishes as the spatial size $L$ increases. 

We want to examine this bound-state pole condition numerically 
in the known bound system. As described previously, it is found
that our simulation in the $Q=4$ case yields an $S$-wave bound state 
as the ground state in the $^{1}S_0$ channel. Thus, we determine 
the phase $\sigma_0$ from an energy level of the 
ground state in the $Q=4$ simulation by using 
L\"uscher's formula~(\ref{Eq.LucsherFormula}).
We first calculate the relative momentum of two particles (electron-positron) from the measured energy level of the ground state $E_{ee}$
by matching with twice the single electron energy $2E_e(p)$.
As we discussed in~Sec.\ref{Sec:3-C2}, we prefer to use the lattice dispersion 
relation~(\ref{Eq:LattDsp}) for a formula of the single electron energy 
$E_e(p)$ in order to avoid lattice discretization errors as much as possible.

\begin{figure}[b]
\begin{center}
\includegraphics[angle=-90,scale=0.3]{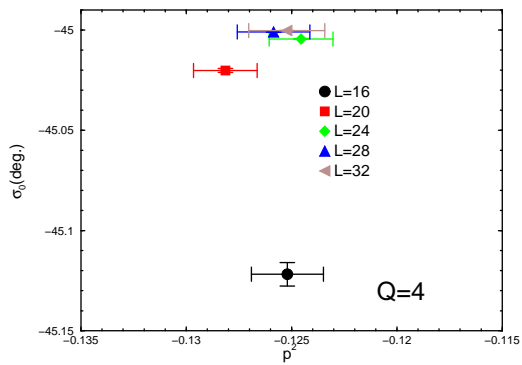}
\end{center}
\caption{
Phase $\sigma_0$ as a function of relative momentum
squared $p^2$ in the $Q=4$ simulation. 
Different symbols represent the values obtained
from simulations with different spatial lattice sizes.}
\label{FIG:Phase_BS}
\end{figure}

In Fig.~\ref{FIG:Phase_BS}, we plot the phase $\sigma_0$ in the $^{1}S_0$ channel 
as a function of $p^2(<0)$. The data plotted appear in Table~\ref{Table:BSphsf_1S0}.
One can easily observe that the phase $\sigma_0$ 
approaches $-45$ deg. ($-\pi/4$) from below as the spatial size $L$ increases. 
Even at the smallest spatial extent $L=16$, the phase $\sigma_0$ is
very close to $-\pi/4$. Needless to say, observed values of $p^2(=-\gamma^2)$, 
which are related to the binding energy of the bound state, are almost
insensitive to the spatial size $L$ within statistical errors.
Thus, we confirm that our observed ``bound state'' in finite volume 
approximately fulfills the pole condition of the $S$-matrix.

\begin{figure}[b]
\begin{center}
\includegraphics[angle=-90,scale=0.3]{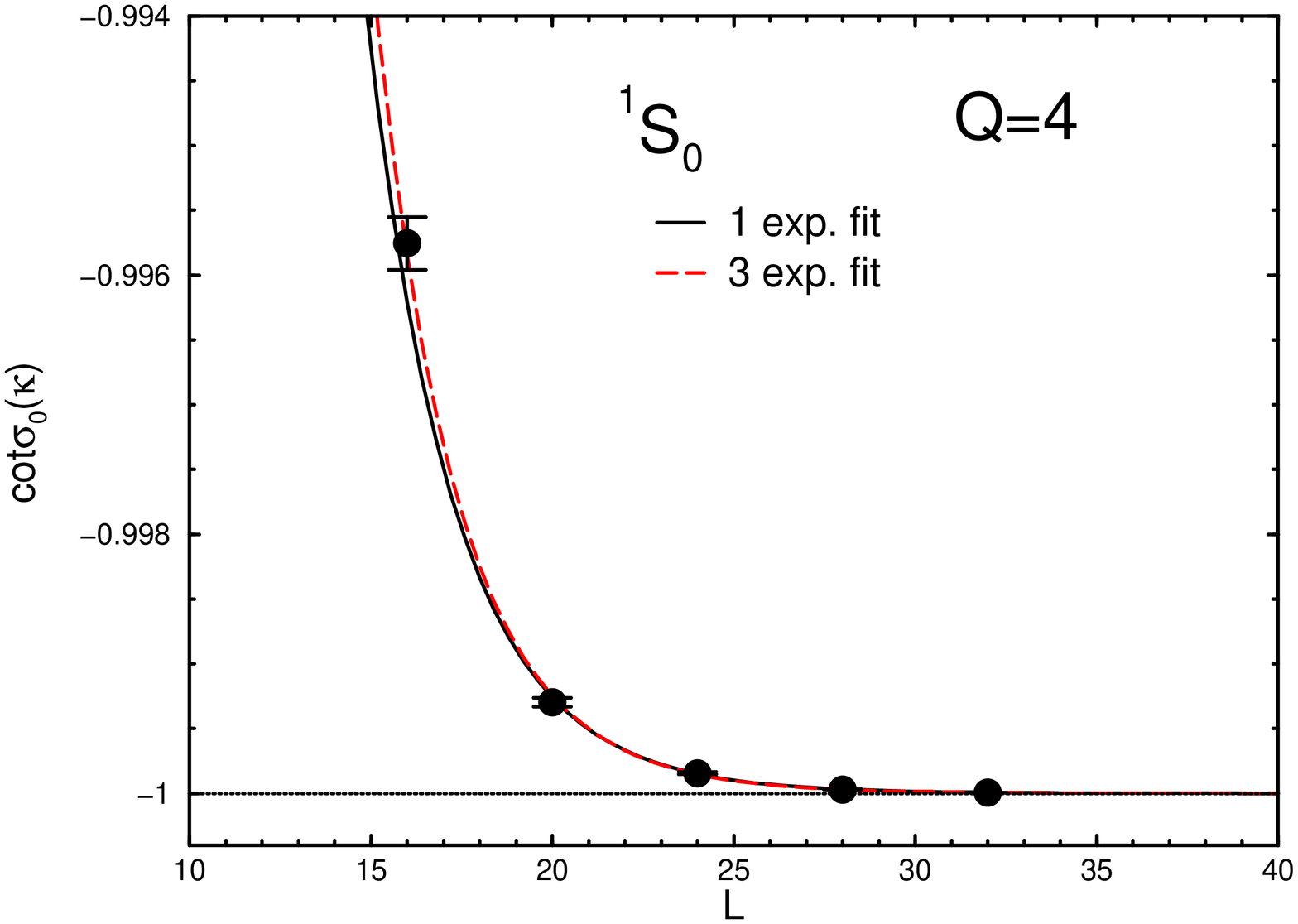}
\end{center}
\caption{$\cot \sigma_0$ in the $^1S_0$ channel for $Q=4$
as a function of the spatial lattice size $L$.
The solid (dashed) curve represents a fitting result using 
Eq.~(\ref{Eq:BScondExp}) with only a leading exponential term 
(three exponential terms).
}
\label{FIG:Ctan_PION_BS}
\end{figure}

A more rigorous way to test for bound-state formation
would be to use an asymptotic formula for the finite volume 
correction to the pole condition as Eq.~(\ref{Eq:BScondExp}).
In Fig.~\ref{FIG:Ctan_PION_BS}, we plot the value of $\cot \sigma_0$
versus the spatial lattice extent $L$ and show two fit results 
using Eq.~(\ref{Eq:BScondExp}) with different numbers of exponential 
terms (one term and three terms). As shown in Table~\ref{Table:BScondFit_1S0},
the optimum number of exponential terms, which yields a convergent result 
of $\gamma$, is about three. However, results with one term and three terms are
quite consistent with each other because of rapid convergence. Then both fit
curves in Fig.~\ref{FIG:Ctan_PION_BS} reproduce all data points very well.

\begin{table}[htdp]
\caption{
Summary of the relative momentum squared $p^2$, the 
phase $\sigma_0$ and $\cot \sigma_0$
measured in the $^1S_0$ channel for $Q=4$ at five different lattice volumes
$L^3 \times 32$.
}
\begin{ruledtabular}
\centering
\begin{tabular}{c|ccccc}
 & 16 & 20 & 24 & 28 & 32 \\\hline
$p^2$ &
$-$0.1252(17) & $-$0.1281(15) & $-$0.1245(15) 
& $-$0.1259(17) & $-$0.1252(18) \\
$\sigma_0$(deg.) & 
$-$45.1218(58) & $-$45.02012(99) & $-$45.00444(26) & $-$45.000860(65) & 
$-$45.000186(17) \\
$\cot\sigma_0$ &
$-$0.99576(20) & $-$0.999298(35) & $-$0.9998449(91) & 
$-$0.9999700(23) & $-$0.99999352(58)
\end{tabular}
\end{ruledtabular}
\label{Table:BSphsf_1S0}
\end{table}
%

%
%
%

%
%
%

\begin{table}[htdp]
\caption{
Fitting results for the bound-state pole condition in the $^1S_0$ channel for $Q=4$ using
Eq.~(\ref{Eq:BScondExp}) with variation of the number of exponential terms.
}
\begin{ruledtabular}
\centering
\begin{tabular}{cccc}
fitting range ($L$) &\# of exp. terms  &$\gamma$ & $\chi^2/{\rm d.o.f.}$ \\\hline
16-32 & 1 & 0.3524(11) & 1.79 \\
            & 2 & 0.3547(11) & 0.86 \\
            & 3 & 0.3549(11) & 0.85 \\
            & 4 & 0.3549(11) & 0.85 
\end{tabular}
\end{ruledtabular}
\label{Table:BScondFit_1S0}
\end{table}
%

\subsection{$e^{-}e^{+}$ elastic scattering phase shifts}
\label{Sec:4-E}

Finally, we evaluate the elastic scattering phase shift of both
the unbound system ($Q=3$) and
the bound system ($Q=4$) using L\"uscher's 
formula~(\ref{Eq.LucsherFormula}).

\subsubsection{Unbound system ($Q=3$)}
\label{Sec:4-E-1}

In the $Q=3$ case, as we described previously, there is no bound state.
The ground state and the first excited state correspond to $n=0$ and
$n=1$ scattering states respectively and those energy levels are 
successfully separated by
the diagonalization method. Then we can measure the 
scattering phase shifts $\delta_0(p)$ at two different kinematical points, which correspond
to the relative momenta of the two particles (electron-positron) $p$ for
both $n=0$ and $n=1$ scattering states. 
However, as for the $n=0$ scattering state, the relative momentum squared is 
negative ($p^2 < 0$) because of the attractive interaction between 
the electron and the positron. Therefore, we only access the phase $\sigma_0$ from
the energy level of the $n=0$ scattering state in this sense. 
However, we consider the effective-range expansion for the scattering phase as 
$p\cot \delta_0(p) = \frac{1}{a_0}+\frac{1}{2}r_0p^2 + {\cal O}(p^4)$ in the vicinity of
zero relative momentum. We then assume this expansion is still valid 
for negative $p^2$. Therefore,
we may translate the phase $\sigma_0$ to
the scattering phase shift $\delta_0$ in the following relation
\be
\lim_{\kappa \rightarrow 0}
\kappa \cot \sigma_0(\kappa)= \lim_{p\rightarrow 0}
p\cot \delta_0(p) = \frac{1}{a_0} \;.
\ee
In other words, we approximately identify the value of $\sigma_0(\kappa)$ at $p^2=-\kappa^2$ to the scattering phase shift $\delta_0(p)$ at $p^2=+\kappa^2$. 
On the other hand, the relative momentum squared 
of the $n=1$ scattering state is definitely positive so that
we directly access the scattering phase shift $\delta_0$ through L\"uscher's formula
without any approximation~\footnote{
Here we remark that there is large systematic uncertainty
to evaluate $p^2$ for the $n=1$ scattering state. This is because resulting $p^2$ 
is rather large so that estimation of $p^2$ strongly depends 
on what type of the dispersion relation was used. }. 

\begin{figure}[b]
\begin{center}
\includegraphics[angle=-90,scale=0.3]{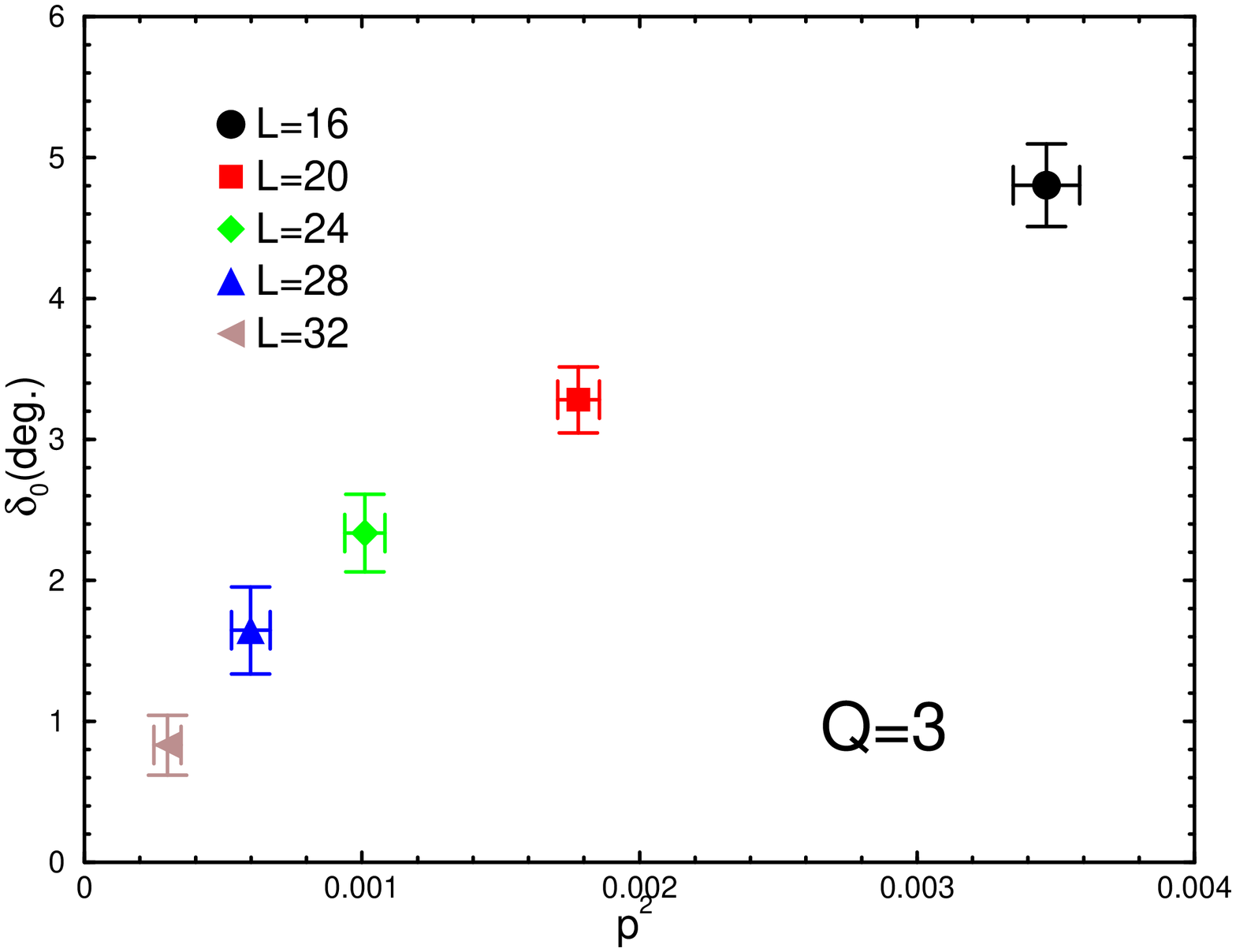}
\includegraphics[angle=-90,scale=0.3]{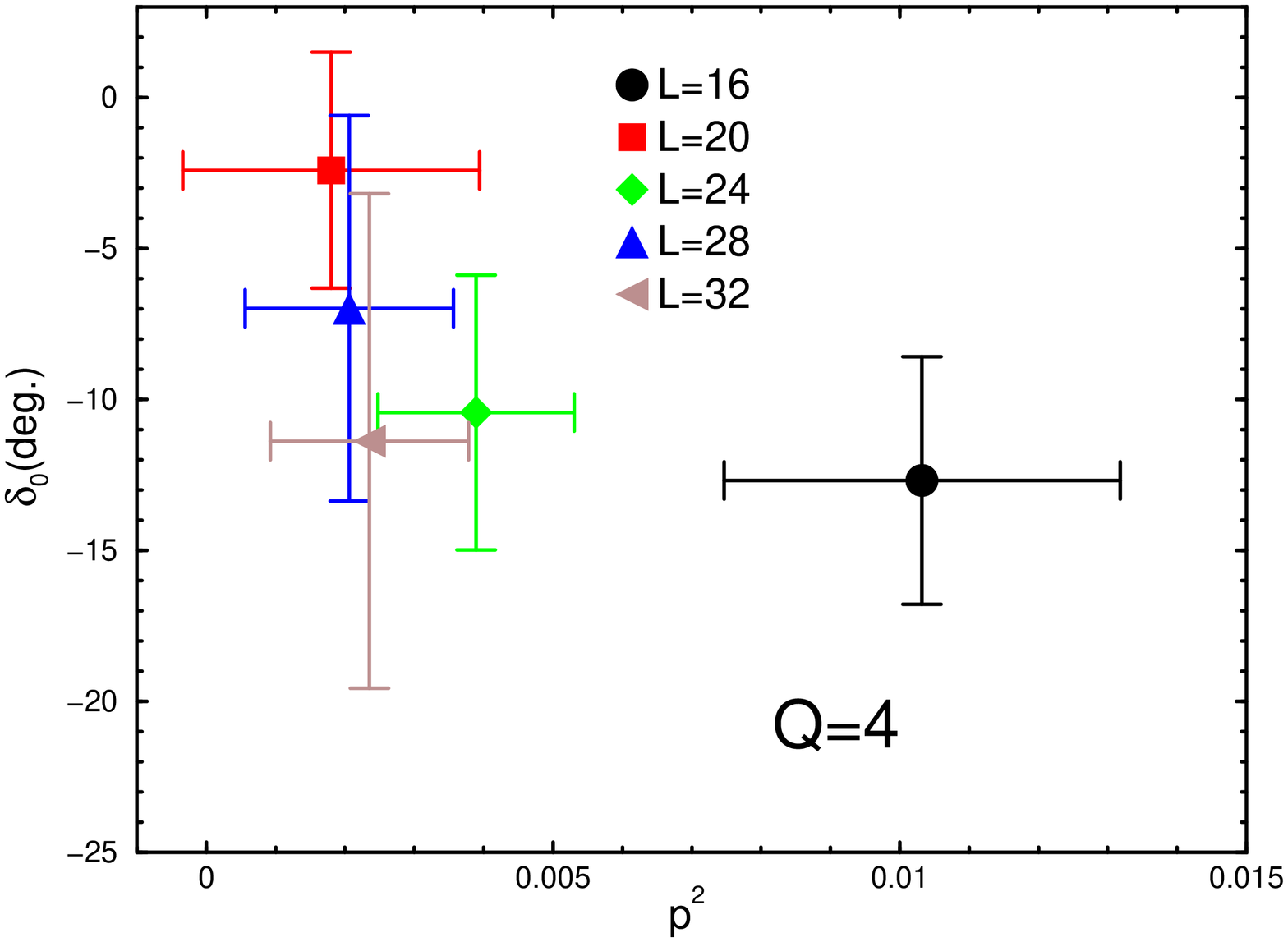}
\end{center}
\caption{Scattering phase shifts $\delta_0$ from the energy level of 
the $n=0$ scattering states (corresponding to 
the ground state for $Q=3$ and the first excited state 
for $Q=4$). The horizontal axis is the squared relative momentum 
of the $e^- e^+$ system.
The left (right) panel is for $Q=3$ ($Q=4$).
Different symbols represent the values obtained
from simulations with different spatial lattice sizes.
Remark that the sign of $\delta_0$ for $Q=4$ is 
opposite to that for $Q=3$.
}
\label{FIG:n0_phase_sft}
\end{figure}

\begin{figure}[htdb]
\begin{center}
\includegraphics[angle=-90,scale=0.3]{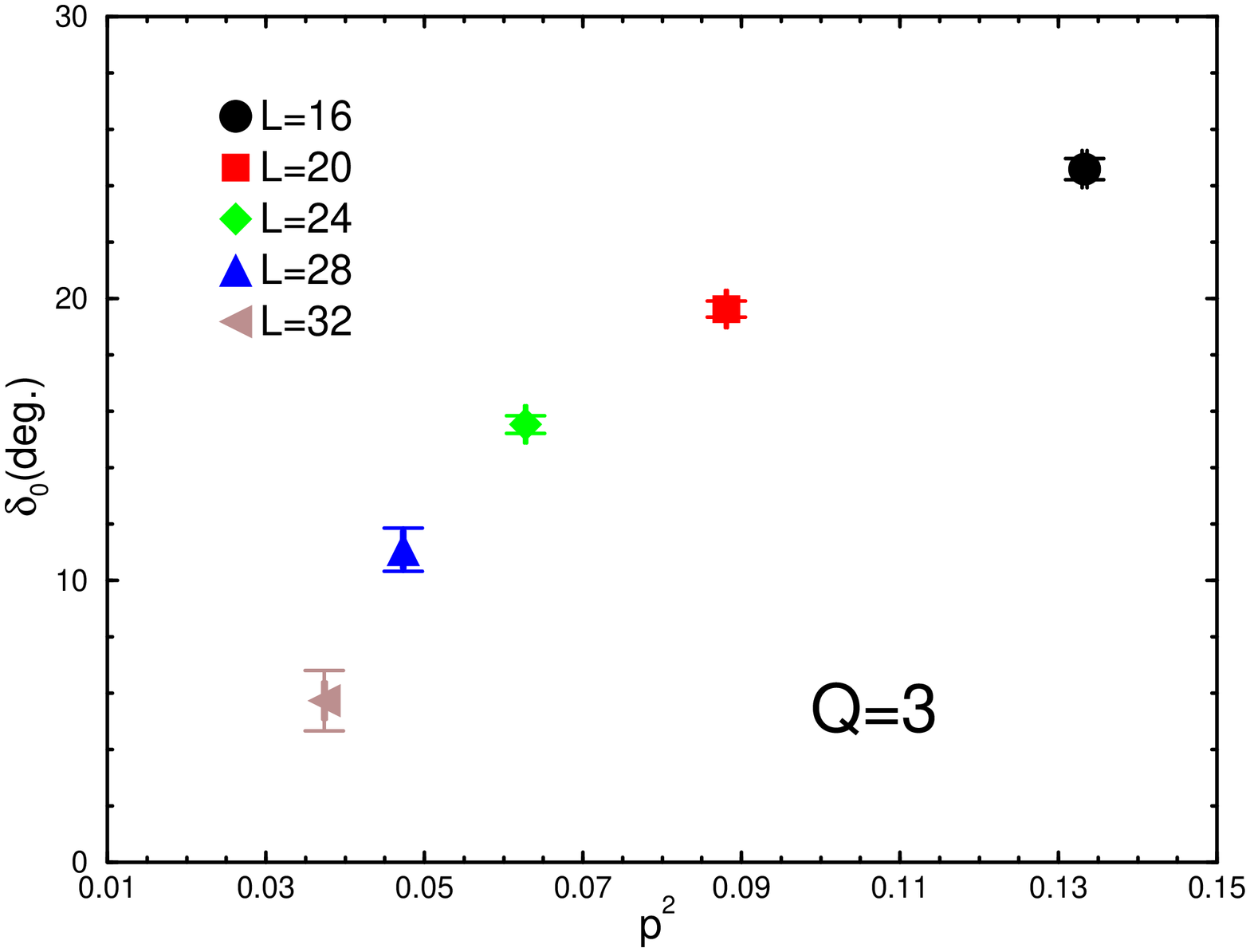}
\includegraphics[angle=-90,scale=0.3]{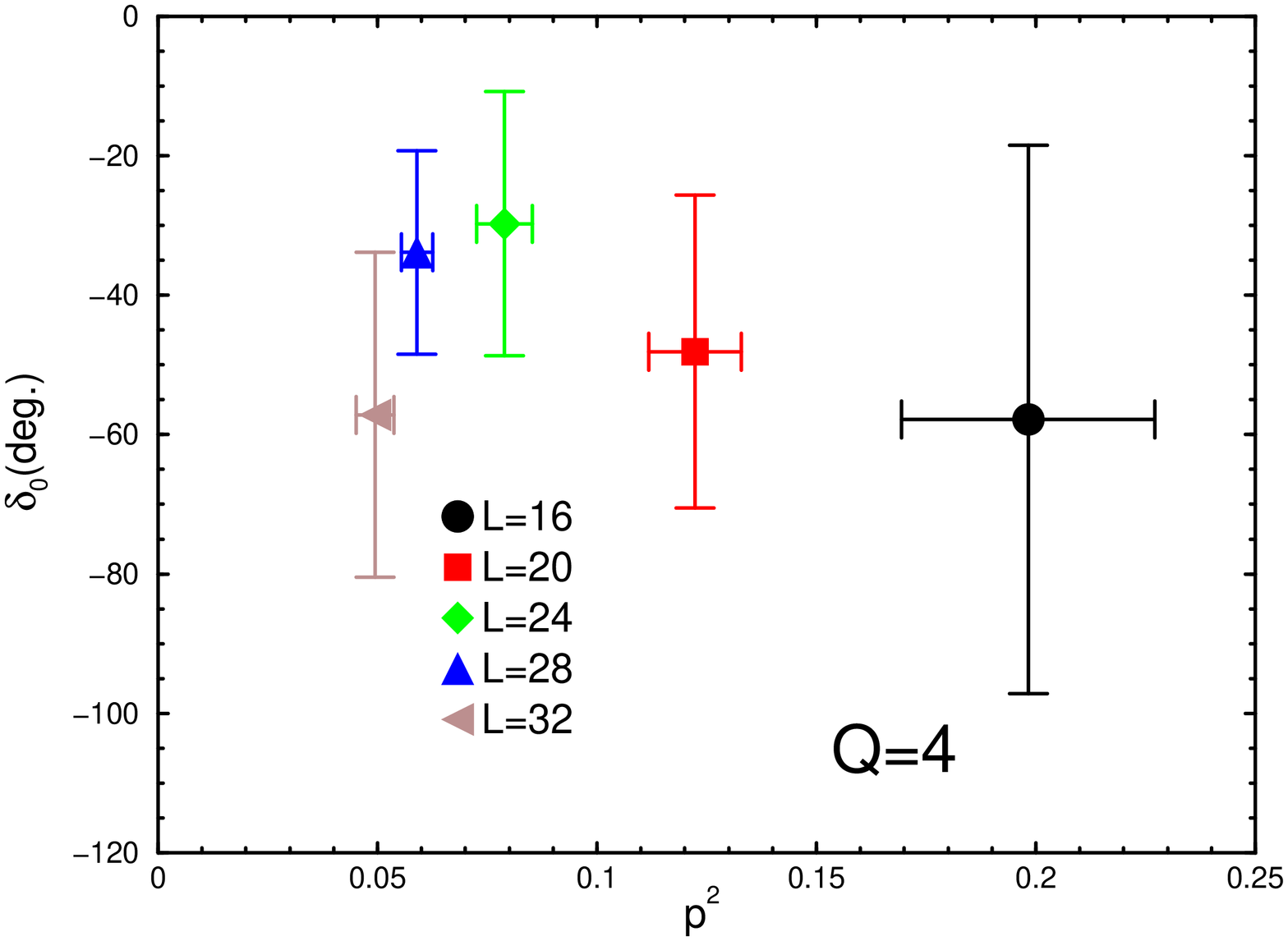}
\end{center}
\caption{Scattering phase shifts $\delta_0$ from the energy level of 
the $n=1$ scattering states (corresponding to 
the first excited state for $Q=3$ and the 
second excited state for $Q=4$). The horizontal axis is the squared 
relative momentum of the $e^- e^+$ system.
The left (right) panel is for $Q=3$ ($Q=4$).
Different symbols represent the values obtained
from simulations with different spatial lattice sizes.
Remark that the sign of $\delta_0$ for $Q=4$ is 
opposite to that for $Q=3$.
}
\label{FIG:n1_phase_sft}
\end{figure}

We plot scattering phase shifts measured from both $n=0$ and 
$n=1$ scattering-state energies in the $Q=3$ case 
as a function of the relative momentum squared 
in the left panel of Figs.~\ref{FIG:n0_phase_sft} and in the left panel
of Figs.~\ref{FIG:n1_phase_sft} following the above descriptions.

%
%
%

%
%
%

\subsubsection{Bound system ($Q=4$)}
\label{Sec:4-E-2}

In the $Q=4$ case, it is found that the ground state corresponds to 
the $S$-wave bound state. As we discussed in Sec.~\ref{Sec:3-B}, 
the first and the second excited states should be $n=0$ and $n=1$ scattering
states, respectively. Although we can not access any information of the scattering
phase shift from the energy level of the ground state, we instead determine 
the scattering phase shift from the energy levels of the first and second excited states. 
In contrast to the purely scattering system without bound states ($Q=3$), 
the relative momentum squared is given as the positive value even 
from the lowest ($n=0$) scattering state, which appears above 
the threshold. Therefore, we can simply evaluate the phase shift $\delta_0$ 
using L\"uscher's formula with measured $p^2$.

In the right panels of both Figs.~\ref{FIG:n0_phase_sft} and 
Figs.~\ref{FIG:n1_phase_sft}, we plot our measured scattering phase shifts
for $Q=4$ versus the relative momentum squared. 
Here, the values of the phase shift $\delta_0$ are 
simply restricted to the interval $(-\frac{\pi}{2}, \frac{\pi}{2}]$. 
Therefore, we observe {\it negative} phase shift $\delta_0$ despite 
an attractive interaction between the electron and the positron. 
Roughly speaking, the phase shift $\delta_0$ monotonically increases as $p^2$ 
decreases and approaches zero toward zero momentum squared. 
It implies that the $S$-wave scattering length $a_0$ is {\it negative} 
for the bound system ($Q=4$).

\subsubsection{Scattering amplitude}
\label{Sec:4-E-3}

Here, we define an $S$-wave scattering amplitude $T(p)$ 
%
%
\be
T(p)=\frac{\tan \delta_0(p)}{p}\frac{E_{ee}}{2},
\ee 
where $E_{ee}$ represents the measured energy of the scattering state.
Analyticity of the scattering amplitude $T(p)$ allows us to consider the 
following fit ans\"atz:
%
%
\be
T(p)=d_0+ d_1 p^2 + d_2 p^4 + d_3 p^6 + d_4 p^8, 
\label{Eq:ScAmpl}
\ee
which is a simple polynomial function in the relative
momentum squared $p^2$. The results of the fit are summarized 
in Table.~\ref{Table:ScAmpFit}.
For $Q=4$, a linear fit with respect to $p^2$ is enough to describe the data with
reasonable $\chi^2/{\rm d.o.f.}$, while the fourth order polynomial fit still
yields large $\chi^2/{\rm d.o.f.}$ in the case of $Q=3$. The latter point will
be discussed before this session is closed. 

\begin{figure}[htdb]
\begin{center}
\includegraphics[angle=-90,scale=0.3]{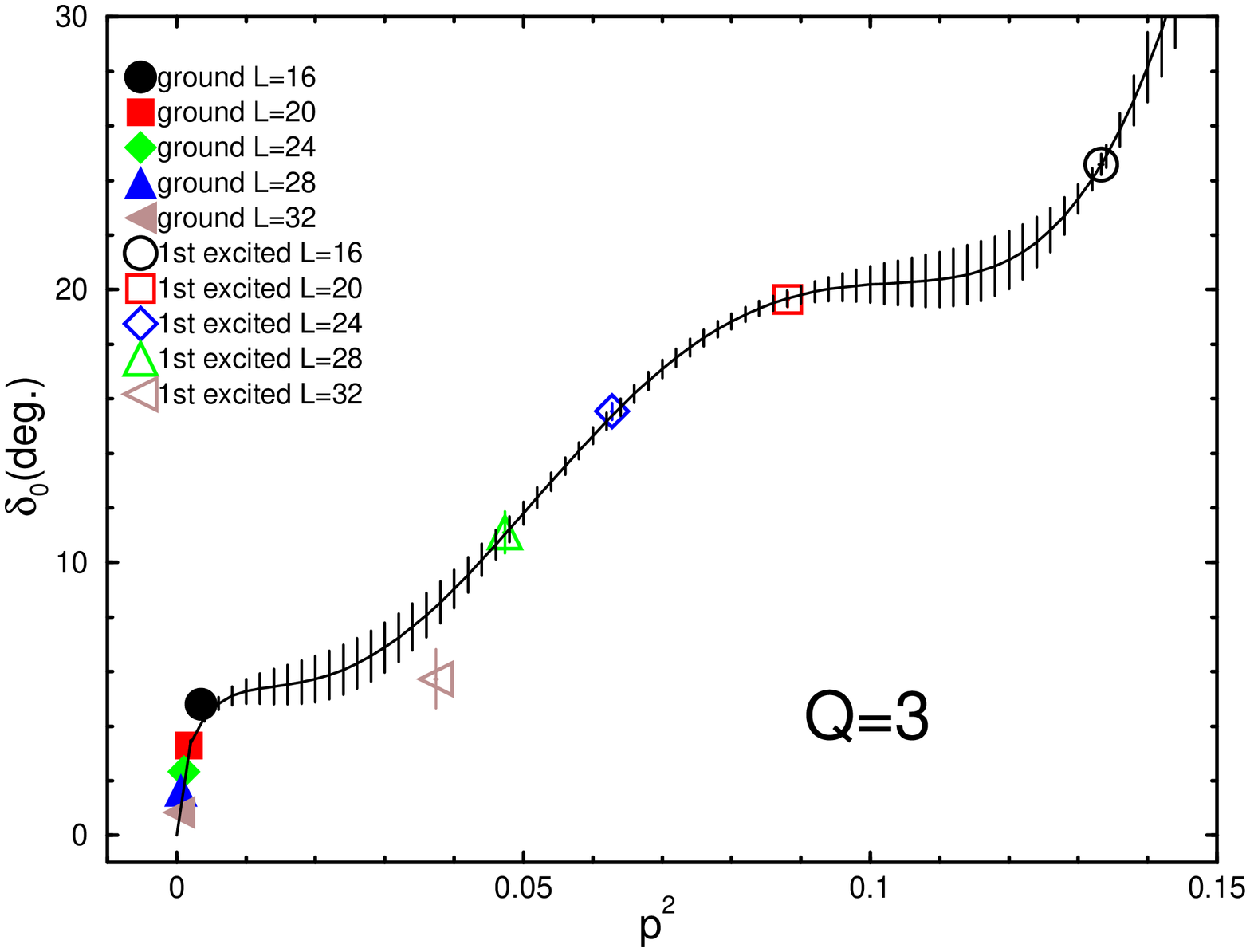}
\includegraphics[angle=-90,scale=0.3]{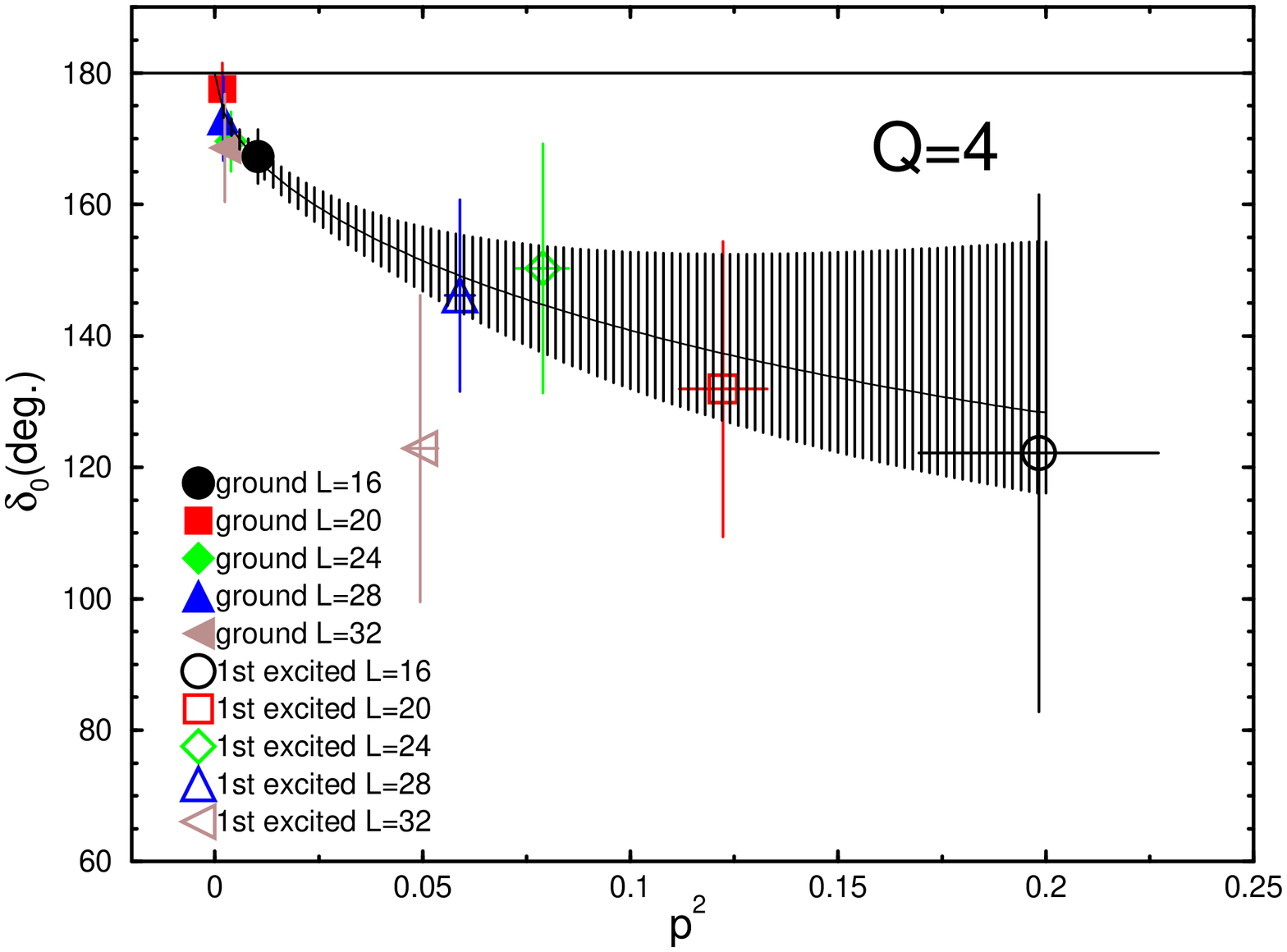}
\end{center}
\caption{All measured scattering phase shift $\delta_0$ 
as a function of the squared relative momentum. As for the $Q=4$ case
(the left panel) where the formation of bound state is observed, 
the scattering phase shifts $\delta_0$ are shifted 
as $\delta_0\rightarrow \delta_0+\pi$ according 
to Levinson's theorem.
Solid curves represent the fitting results with the 
band of their errors.
}
\label{FIG:all_phase_sft}
\end{figure}

We then obtain a global $p^2$-dependence
of the phase shift $\delta_0$ in the measured region of $p^2$, which is
deduced from fitted results of the scattering amplitude.
We show all measured phase shifts $\delta_0$,
which are obtained from the energy levels of 
both $n=0$ and $n=1$ scattering states, 
in Figs.~\ref{FIG:all_phase_sft}.
Solid curves represent inferred $p^2$-dependence of the phase shift
with the band of their errors.

In the right panel ($Q=4$), we take into account the modulo-$\pi$ ambiguity in 
determination of the phase shift $\delta_0$ because of the bound system
and raise the phase shift by an additional $\pi$ in order 
to fulfill Levinson's theorem. That is why the phase shift data starts 
from $\pi$ and monotonically decreases as $p^2$ increases. 
All data are well covered with rather wide bands of error
associated with the global fit.

On the other hand, in the left panel ($Q=3$), two data sets determined
from energy levels of $n=0$ and $n=1$ scattering states
seem not to be smoothly connected with each other due to the lower data points 
from the $n=1$ scattering state at $L=28$ and 32. 
We remark that although statistical errors on all points
are rather small, a hidden and large systematic error stems from an 
order ${\cal O}(a)$ lattice artifact in the determination of $p^2$.
As we discussed in Sec.~\ref{Sec:3-C2}, we have used the lattice dispersion
relation in the analysis of the scattering phase shift. 
The continuum-type dispersion relation yields smaller estimations of $p^2$ than
those obtained from the lattice dispersion relation. These differences are 
far beyond statistical errors, especially for $p^2$ obtained from the $n=1$ energy level
in the $Q=3$ case.
Furthermore, discrepancies are largely enhanced in determination of the 
scattering phase shift through the L\"uscher finite formula. 
The scattering phase shift from the $n=1$ energy level for $Q=3$ typically 
increases by about a factor of two, if the continuum-type dispersion relation is utilized in the 
whole analysis.

At the low-energy limit, the scattering amplitude becomes
%
%
\be
\lim_{p \rightarrow 0} T(p) = a_0 M_e .
\ee
Therefore, the fitting parameter $d_0$ in Eq.~(\ref{Eq:ScAmpl}) is associated with 
the scattering length $a_0$. We then obtain
the scattering lengths as $a_0=1.46(5)$ for $Q=3$ and $-2.28(40)$ for $Q=4$ in 
lattice units, which are much smaller than our utilized lattice sizes ($L\ge12$).
Needless to say, the sign of the scattering length for $Q=4$ is opposite to that for
$Q=3$ due to formation of one bound state in the case of $Q=4$.

\begin{table}[htdp]
\caption{
Fitting results for the scattering amplitude for $Q=3$ and $Q=4$ 
using Eq.~(\ref{Eq:ScAmpl}).
}
\begin{ruledtabular}
\centering
\begin{tabular}{c|cccccc}
charge $Q$ & $d_0$ & $d_1$ & $d_2$ & $d_3$ & $d_4$ & $\chi^2/$d.o.f.\\\hline
3 & 0.697(26) & $-$33.8(7.4) & 102(25)$\times 10$ & $-$103(28)$\times 10^2$
  & 34(10)$\times 10^3$ & 7.26 \\
4 & $-$1.15(20) & $-$4.6(7.3) & --- & --- & --- & 0.52 \\
\end{tabular}
\end{ruledtabular}
\label{Table:ScAmpFit}
\end{table}
%

%
%
%

\section{Numerical results in the $^{3}S_1$ channel}
\label{Sec:5}

\subsection{Low-lying spectra in the $Q=4$ case}
\label{Sec:5-A}

The $^{3}S_1$ spectroscopy has been done in exactly the same way as the $^{1}S_0$ 
case by using the bilinear vector operator $\overline{\Psi}_x\gamma_\mu \Psi_x$. 
As for the Lorentz indices, we take an average over the spatial indices so as to 
gain possible reduction of statistical errors.  After we perform the diagonalization method with
the $3\times 3$ matrix correlator constructed with three operators in Eq.(\ref{Eq:ThreeOps}),
we get the energy spectra of both the ground state and the first excited state.

In the $Q=4$ case, we have concluded that one bound state is formed 
in the $^{1}S_0$ channel as described in the previous section. 
The binding energy $B=|M_{\rm bs}-2M_e|$ is rather large as 
$B \approx M_{e}/2$. The observed bound state should be 
a ``tightly bound state" rather than a ``loosely bound state".
On the other hand, the mass of the $^{3}S_1$ bound state is naturally 
expected to be higher than the $^{1}S_0$ bound state due to the 
hyperfine-splitting interaction. Indeed, we observe that the ground state in the $^{3}S_1$ 
channel is much closer to the threshold energy as $|\Delta E| \approx M_{e}/25$.
Although the energy level of the ground state is too near the threshold to be 
simply identified as a bound state, we may expect that the $^{3}S_1$
ground state is a near-threshold bound state or a loosely bound state. 
Needless to say, to draw a solid conclusion, we need more 
rigorous signatures of bound-state formation in the $^{3}S_1$ channel.

\begin{figure}[b]
\begin{center}
\includegraphics[angle=-90,scale=0.3]{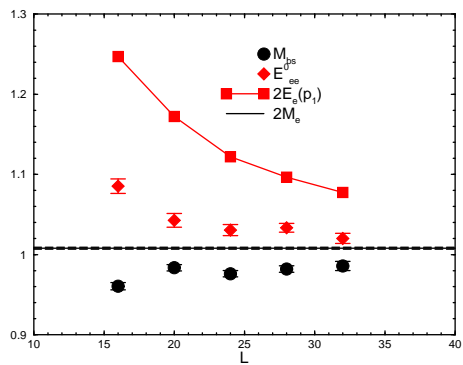}
\end{center}
\caption{Energies of the ground state and the first excited state in
the $^{3}S_1$ channel of the $e^- e^+$ system as a function of 
spatial lattice size for $Q=4$. The horizontal line represents
the threshold energy determined by $2M_e$. Full circles
and full diamonds are measured energies for the ground state and 
the first excited state respectively.
A solid curve with full squares shows twice of the single electron energy
with non-zero smallest momentum $p_1=2\pi/L$.
}
\label{FIG:3S1state}
\end{figure}

We employ the diagonalization method to separate the first excited state from the
ground state. Fig.~\ref{FIG:3S1state} shows $L$-dependence of energies of the ground state 
and the first excited state in the $^{3}S_1$ channel for $Q=4$.
The horizontal axis is the spatial size $L$ and the vertical axis is
the energy of the ground state (full circles) or the first excited state (full diamonds).
The horizontal lines represent the threshold energy of the $e^- e^+$ system
together with the 1 standard deviation, which is evaluated as twice the 
measured electron mass. Although it seems that the ground state has no appreciable 
finite-size effect for $L$ larger than 20, the $^{3}S_1$ ground state
lies too close to the threshold energy to be assured of bound-state formation.

As shown in Sec.~\ref{Sec:4-C}, the distinctive signature of bound states is
given by an information of the excited state spectra: if a bound state is formed, 
the lowest ($n=0$) scattering state could appear just above the threshold ($2M_e$), 
but far from the $n=1$ energy level of non-interacting two-electron system ($2E_e(p_1)$).
Indeed, we observe that the first excited state appears just above the threshold
and its energy rapidly approaches the threshold as spatial size $L$ increases. 
The first excited state can be clearly distinguished from the $n=1$ scattering state.
Of course, it indicates that the ground state should not be the lowest scattering state.
Thus, we can conclude: the $^{3}S_1$ ground state should be the $S$-wave bound state, 
of which formation clearly induces the sign of the scattering length to change. 
Therefore, the lowest ($n=0$) scattering state approaches the threshold from above,
the same as the repulsive system in the attractive channel. 
This result shows that our proposal could be quite promising for identifying 
a near-threshold bound state or a loosely bound state such as 
a hadronic molecular state in a finite box on the lattice.

%
%
%

\subsection{Bound-state pole condition}
\label{Sec:5-B}

Next, we evaluate the phase $\sigma_0$ from the energy level of the ground state 
through the phase-shift formula~(\ref{Eq.LucsherFormula}) as we did
in Sec.~\ref{Sec:4-C-2}. All results measured at five different lattice 
volumes $L^3 \times 32$ are tabulated together with results of $p^2$ and
$\cot \sigma_0$ in Table~\ref{Table:BSphsf_3S1}.  Indeed, we observe that the phase $\sigma_0$ gradually approaches $-45$ deg. ($-\pi/4$) as spatial lattice 
extent $L$ increases. However, $\sigma_0$ is not really close to $-45$ deg. 
even at the largest volume ($L=32$), in comparison to $\sigma_0$ from the smallest volume ($L=16$) in the $^1S_0$ channel. In this sense, it is hard to judge how large of a lattice size 
is enough to deal with the asymptotic solution of the bound state even in finite volume.
Thus, we should examine the $L$-dependence of the specific quantity, $\cot \sigma_0$,
by reference to Eq.~(\ref{Eq:BScondExp}), where the finite volume
corrections on the bound-state pole condition are theoretically predicted.

\begin{figure}[b]
\begin{center}
\includegraphics[angle=-90,scale=0.3]{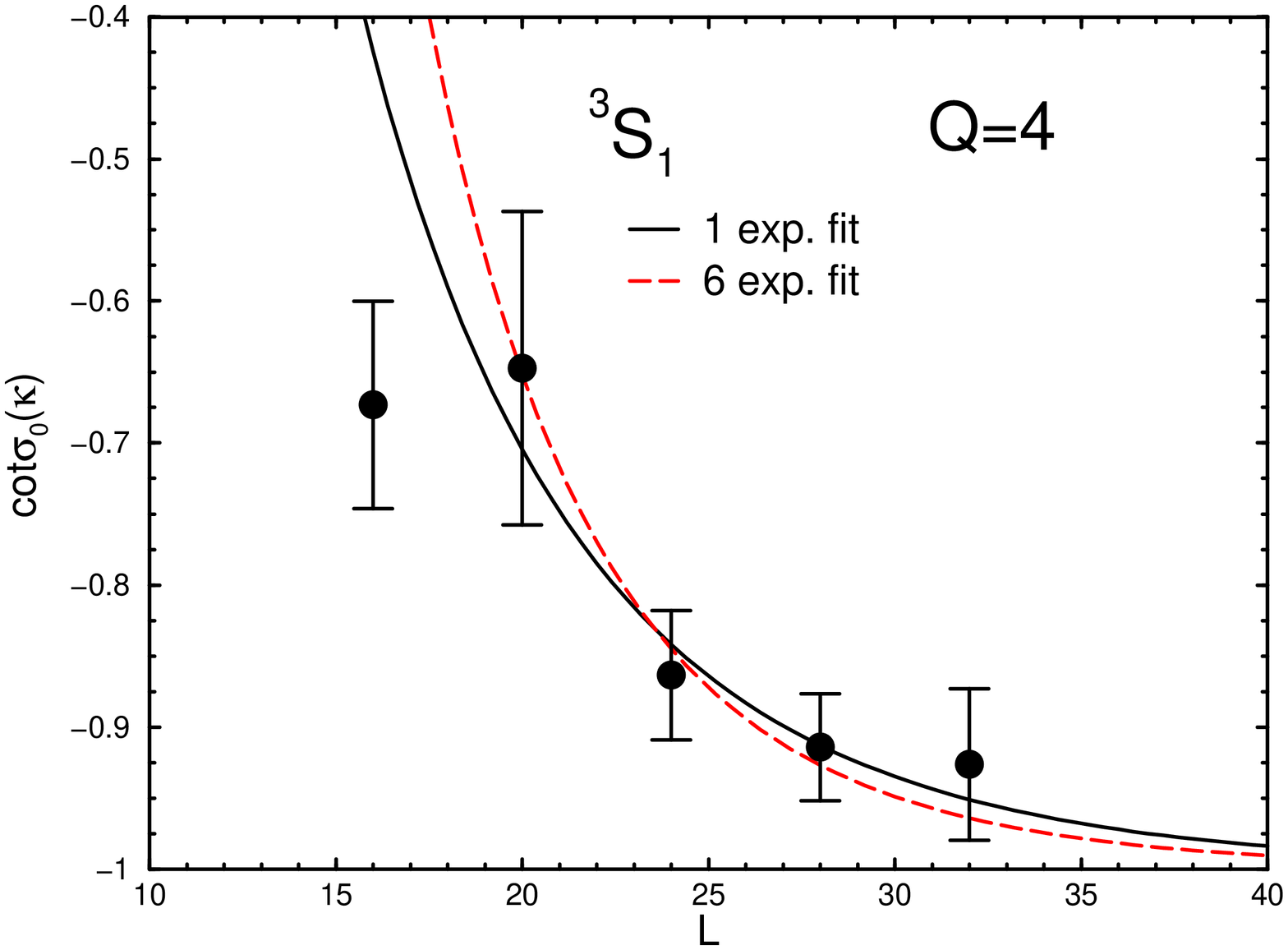}
\end{center}
\caption{$\cot \sigma_0$ in the $^3S_1$ channel for $Q=4$
as a function of the spatial lattice size $L$.
The solid (dashed) curve represents a fitting result using 
Eq.~(\ref{Eq:BScondExp}) with only a leading exponential term 
(six exponential terms).}
\label{FIG:Ctan_RHO_BS}
\end{figure}

As shown in Fig.~\ref{FIG:Ctan_RHO_BS}, the values of $\cot \sigma_0$
are plotted as a function of spatial lattice extent $L$. 
Full circles are measured
value at five different lattice volumes. The solid and dashed curves represent
fit results with a single leading exponential term and six exponential terms 
in Eq.~(\ref{Eq:BScondExp}). The four data points in the region $20\le L \le 32$
are used for those fits. The fitting with the six exponential terms yields
a convergent result of $\gamma$ as shown in Table~\ref{Table:BScondFit_3S1}.
Either fit curves in Fig.~\ref{FIG:Ctan_RHO_BS} reproduce all data points except
for data at the smallest $L$. Indeed, the resulting $\chi^2/{\rm d.o.f.}$ is no longer 
reasonable as $\chi^2/{\rm d.o.f.}\approx 3$ if the data point at $L=16$ is used. Therefore, the ground state at least for $L\ge 20$ can be 
identified as a bound state without ambiguity.

\begin{table}[htdp]
\caption{
Summary of the relative momentum squared $p^2$, the 
phase $\sigma_0$ and $\cot \sigma_0$
measured in the $^3S_1$ channel for $Q=4$ at five different lattice volumes
$L^3 \times 32$.}
\begin{ruledtabular}
\centering
\begin{tabular}{c|ccccc}
 & 16 & 20 & 24 & 28 & 32 \\\hline
$p^2$ &
$-$0.0255(26) & $-$0.0158(23) & $-$0.0167(23) & 
$-$0.0148(26) & $-$0.0120(34) \\
$\sigma_0$(deg.) & 
$-$56.1(2.9) & $-$57.1(4.5) & $-$49.2(1.5) & $-$47.6(1.2) & $-$47.2(1.6) \\
$\cot\sigma_0$ &
$-$0.673(73) & $-$0.65(11) & $-$0.864(45) & $-$0.914(38) & $-$0.926(53) \\
\end{tabular}
\end{ruledtabular}
\label{Table:BSphsf_3S1}
\end{table}
%

%
%
%

\begin{table}[htdp]
\caption{Fitting results for the bound-state pole condition in the $^3S_1$ channel for $Q=4$ using
Eq.~(\ref{Eq:BScondExp}) with variation of the number of exponential terms.
}
\begin{ruledtabular}
\centering
\begin{tabular}{cccc}
fitting range ($L$) &\# of exp. terms  &$\gamma$ & $\chi^2/{\rm d.o.f.}$ \\\hline
20-32 & 1 & 0.1109(60) & 0.24 \\
            & 2 & 0.1218(57) & 0.24 \\
            & 3 & 0.1236(56) & 0.25 \\
            & 4 & 0.1242(55) & 0.25 \\
            & 5 & 0.1251(55) & 0.26 \\
            & 6 & 0.1256(54) & 0.27 \\
            & 7 & 0.1256(54) & 0.27 
\end{tabular}
\end{ruledtabular}
\label{Table:BScondFit_3S1}
\end{table}

\section{Summary and conclusion}
\label{Sec:6}

In this paper, we have discussed signatures of bound-state formation
in finite volume via L\" uscher finite size method.
Assuming that the phase-shift formula inherits all aspects of
the quantum scattering theory, we can propose a novel approach 
to distinguish a ``loosely bound state" from the lowest scattering state, which 
is located below the threshold in finite volume in the case of attractive 
two-particle interaction.
According to the quantum scattering theory, the $S$-wave scattering length 
is positive ($a_0>0$) in the attractive channel, if the attraction is not strong 
enough to give rise to a bound state. However, the sign of the scattering 
length turns out to be opposite ($a_0<0$) once the bound state is formed. 
This fact provides us a distinctive identification of a loosely bound state 
even in finite volume through the observation of the lowest scattering state 
that is above the threshold. We also reconsider the bound-state pole condition
in finite volume, based on the phase-shift formula in the L\"uscher finite
size method. We find that the bound-state pole condition is fulfilled
only in the infinite volume limit, but its modification by finite size corrections
is exponentially suppressed by the spatial lattice size $L$. 

To check the above theoretical considerations, we have performed
numerical simulations to calculate the positronium spectrum in compact 
scalar QED, where the short-range interaction between an electron and 
a positron is realized in the Higgs phase. We introduce the fictitious
$Q$-charged electron to control the strength of this interparticle force and
then can adjust the charge $Q$ to give rise to the $S$-wave bound states such as
$^1S_0$ and $^3S_1$ positronium. We choose two parameter sets 
$(Q, \kappa)$ that lead to an unbound $e^- e^+$ system ($Q=3$) and a bound 
$e^- e^+$ system ($Q=4$) at approximately the same mass of a single electron. 
We observe the following signatures of the bound-state formation, some of which
are related to our theoretical proposals, in our numerical simulations. 

\bi
 \item 
 The lowest scattering state has better overlap with the wall-wall correlator 
 than the point-point correlator. This tendency is inverted in the case of the bound state.

\item The sign of the $S$-wave scattering length turns out to be opposite ({\it repulsive-like})
even in the attractive channel, once the bound state is formed.

\item In the bound system, the phase $\sigma_0$, which is related to the 
 scattering phase $\delta_0(p)$ and analytically continued into the complex $p$-plane, 
 is near $-45$ deg. ($-\pi/4$) which is associated with the pole condition of 
 the $S$-matrix.
 
 \item The deviation from the pole condition, $\cot \sigma_0=-1$,
 in finite volume is well described by a finite series of exponentially
 convergent terms with respect to the spatial extent $L$ scaled by 
 the binding momentum $\gamma$. 
\ei

In particular, we regard the second point, the bound-state formation induces
the sign of the scattering length to be changed, as crucially important for
identifying a ``loosely bound state". This is because one can distinguish
it from the lowest scattering state even in a single simulation at fixed $L$
through determination of whether the second lowest energy state appears
just above the threshold or near the $n=1$ energy level of 
non-interacting two-particle system.
We also emphasis that L\"uscher's phase-shift formula properly
reflects one of the most essential features of the quantum scattering theory,  namely Levinson's theorem.

\begin{acknowledgments}
We would like to thank T. Blum for helpful suggestions and his careful reading of the manuscript. 
We also thank RIKEN, Brookhaven National Laboratory and the U.S. DOE for providing the 
facilities essential for the completion of this work. 
The results of calculations were performed by using of RIKEN Super Combined Cluster (RSCC).
Finally, we are grateful to all the members of the RIKEN BNL Research
Center for their warm hospitality during our residence at Brookhaven National Laboratory.

\end{acknowledgments}

\section*{Appendix A: Sensitivity of mass spectra to spatial boundary conditions}
\label{Apx-A}

Kinematics of two-particle states on the lattice should be sensitive to choice 
of the spatial boundary condition. In many literatures, this particular 
point is often discussed and sometimes applied to explore hadronic or non-leptonic decay
processes~\cite{{DeGrand:1990ip},{Kim:2002np}} or to search 
exotic hadrons~\cite{{Sasaki:2003gi}, {Csikor:2003ng}}
on the lattice. The main point is that the total energy of two-particle states, 
which is roughly estimated by a sum of the energy of non-interacting two particles, 
depends on the spatial size $L$ unless the relative momentum of two particles is zero. 
Of course, this is because all momenta on the lattice are discretized in units of $2\pi/L$.
Here, we have considered the $e^{-}e^{+}$ system. The total energy of 
electron-positron states is approximately estimated by using the naive relativistic dispersion 
relation in the following.
%
%
\be
E_{ee}^n\sim2E_{e}(p_{n})=2\sqrt{M_e^2+{\bf p}_n^2}.
\ee
The discrete momenta of a single electron are obtained as ${\bf p}_n=\frac{2\pi}{L}{\bf n}$ for
the periodic boundary condition (P.B.C.) and ${\bf p}_n=\frac{\pi}{L}(2{\bf n}+1)$
for the anti-periodic boundary condition (A.P.B.C.).
In the anti-periodic boundary condition, zero relative-momentum is not kinematically
allowed, so that the lowest energy of two-particle scattering states is expected
to be very sensitive to the spatial lattice size. In other words, 
different types of spatial boundary conditions (periodic or anti-periodic)
exhibit different energy levels of the two-particle 
scattering states even at the fixed spatial size, while a mass of $e^-e^+$ bound 
states (positronium states) should be insensitive to the spatial boundary 
condition for the electron fields~\footnote{
This is because the positronium states, of which a single particle two-point correlator 
contains even numbers of electron propagators, are totally subjected to the
periodic boundary condition in either periodic or anti-periodic boundary 
conditions for the electron fields.}. 
For an example, in the $n=0$ case, we obtain 
\be
E_{ee}^0\sim2E_{e}(p_{0})=
\left\{
\begin{array}{ll}
2 M_e & \mbox{(P.B.C. for all spatial directions)},\\
\\
2\sqrt{M_e^2+3\cdot \left(\frac{\pi}{L}\right)^2} & \mbox{(A.P.B.C. for all spatial directions)},
\end{array}
\right.
\ee
which imply an inequality $E^{0}_{ee}({\rm P.B.C.})<E^{0}_{ee}({\rm A.P.B.C.})$.
In general, we can expect that $E^{n}_{ee}({\rm P.B.C.})<E^{n}_{ee}({\rm A.P.B.C.})$ 
is always fulfilled.

Recently, such sensitivity of spatial boundary condition is often utilized to distinguish 
between two-particle scattering states and a single-particle state 
(a bound state or a resonance state)~\cite{{Ishii:2004qe},{Iida:2006mv}}. 
However, there is no rigorous test of 
whether this approach is adequate for such purpose so far.
In this subsection, we examine this approach in our simulated $e^{-}e^{+}$ system.
 
We use the following operators under  {\it the anti-periodic spatial boundary condition} 
for electron fields:
%
%
\bea
\Omega_P(t)&=&\frac{1}{L^3}\sum_{\bf x}{\overline \Psi}({\bf x}, t)\gamma_5 \Psi({\bf x},t), \\
\Omega_{M_0}(t)&=&\frac{1}{L^6}\sum_{{\bf x},{\bf y}}{\overline \Psi}({\bf y}, t)\gamma_5 \Psi({\bf x},t)e^{i{\bf p}_0\cdot({\bf x}-{\bf y}) }, \\
\Omega_{M_1}(t)&=&\frac{1}{L^6}\sum_{{\bf x},{\bf y}}{\overline \Psi}({\bf y}, t)\gamma_5 \Psi({\bf x},t)e^{i{\bf p}_1\cdot({\bf x}-{\bf y}) },
\eea
where ${\bf p}_0=\frac{\pi}{L}(1,1,1)$ and ${\bf p}_1=\frac{\pi}{L}(3,1,1)$.
The first operator is a simple local-type operator. 
The second and third operators
project both the electron and the positron on 
non-zero lowest momentum ($|{\bf p}_0|=\frac{\sqrt{3}\pi}{L}$)
and non-zero second lowest momentum
($|{\bf p}_1|=\frac{\sqrt{11}\pi}{L}$), respectively.
We can expect that $n=0$ and $n=1$ scattering states have
strong overlap with $\Omega_{M_0}$ and $\Omega_{M_1}$, while the bound state
has the better overlap with $\Omega_{P}$ than $\Omega_{M_0}$ and $\Omega_{M_1}$.

\begin{figure}[b]
\begin{center}
\includegraphics[angle=-90,scale=0.3]{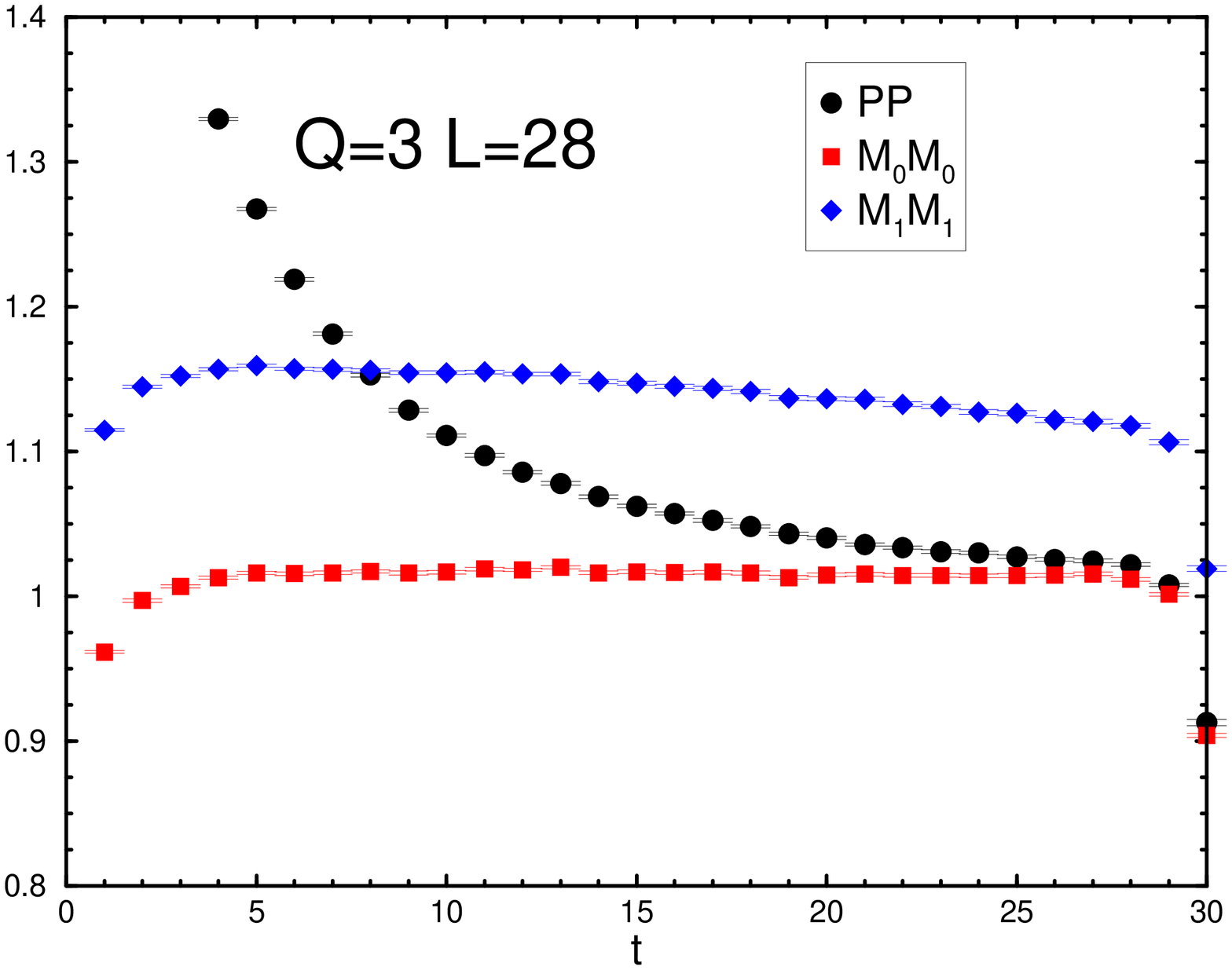}
\includegraphics[angle=-90,scale=0.3]{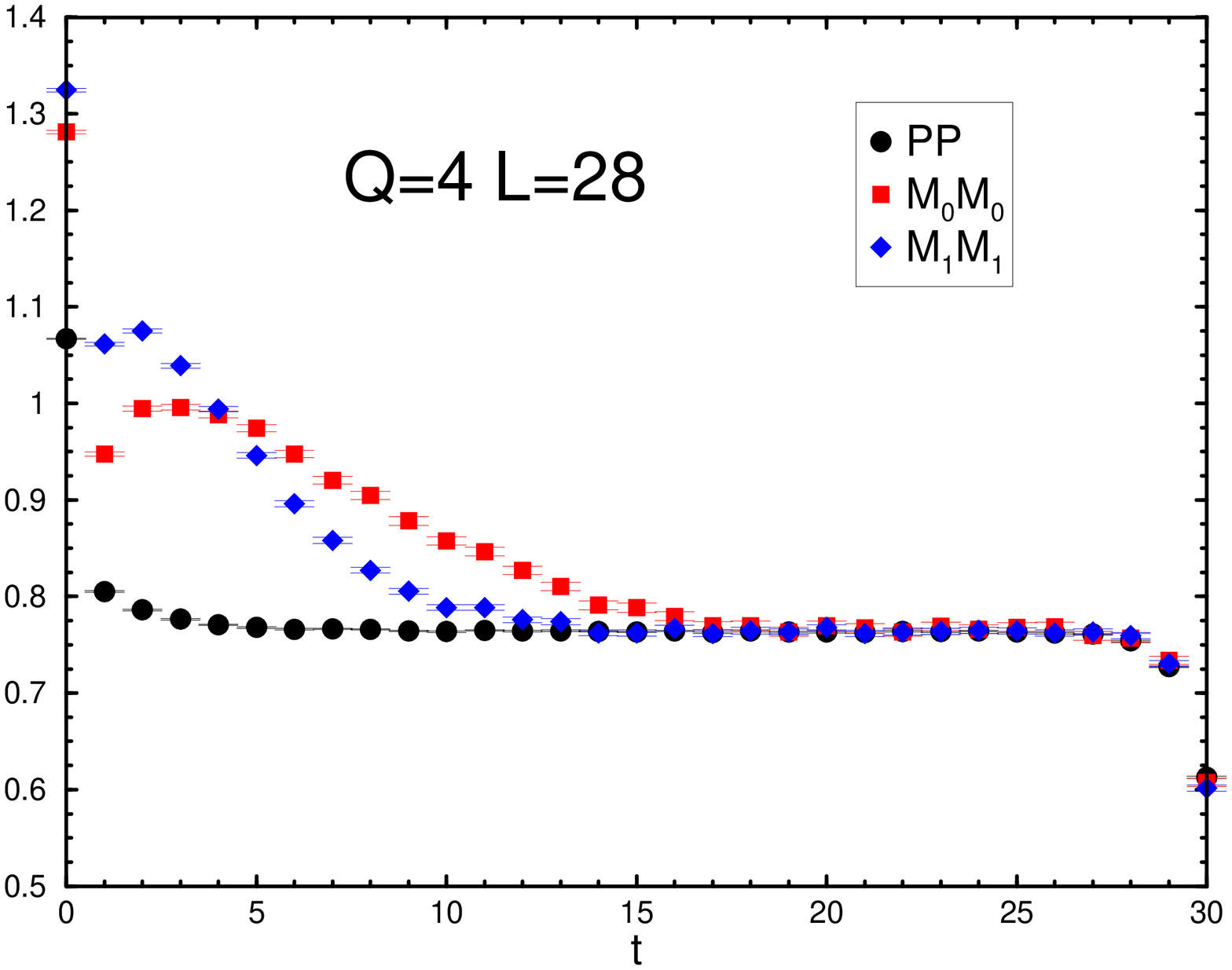}
\end{center}
\caption{
The effective masses in the $^{1}S_{0}$ channel on the lattice with
$L=28$ as a function of the time-slice $t$ in lattice units
The anti-periodic boundary condition is imposed in the spatial direction.
The left (right) panel is for $Q=3$ ($Q=4$) electron fields.
Full circles, full squares and full diamonds are obtained from $PP$, $M_0 M_0$
and $M_1 M_1$ correlators respectively. }
\label{FIG:OpDepAPBC}
\end{figure}

Figs.~\ref{FIG:OpDepAPBC} show the effective mass plots of the $PP$, $M_0 M_0$ and
$M_1 M_1$ correlators in simulations at $L=28$ for $Q=3$ (the left panel) and $Q=4$
(the right panel) in the $^{1}S_0$ channel.
There is a similarity between Figs.~\ref{FIG:OpDep} (P.B.C.) and Figs.~\ref{FIG:OpDepAPBC} (A.P.B.C.). Very clear plateaus are given by the $M_0 M_0$ correlator
in the $Q=3$ case and the $PP$ correlator in the $Q=4$ case in Figs.~\ref{FIG:OpDepAPBC},
while the same quality shows up for the $WW$ correlator in the $Q=3$ case and
the $PP$ correlator in the $Q=4$ case in Figs.~\ref{FIG:OpDep}. In either P.B.C. and
A.P.B.C. cases, the $PP$ correlator strongly overlap with the $Q=4$ ground state, 
which has already been identified as the bound state in Sec.~\ref{Sec:4}. 
Both $WW$ and $M_0 M_0$ correlators are expected to have large overlap 
with the lowest ($n=0$) scattering state
under each spatial boundary condition. A main difference between 
Figs.~\ref{FIG:OpDep} (P.B.C.) and Figs.~\ref{FIG:OpDepAPBC} (A.P.B.C.) is that 
the $PP$ ($M_1 M_1$) correlator for $Q=3$ ($Q=4$) in the A.P.B.C. approaches the plateau much faster than the P.B.C. cases. This is simply because $E^n_{e}({\rm A.P.B.C.})$ is larger than $E^n_{e}({\rm P.B.C.})$ and then propagations of non-ground state can die out more quickly in the A.P.B.C. case than the P.B.C. case. Indeed, the $M_1 M_1$ correlator
approaches the plateau faster than the $M_0 M_0$ correlator in the left panel ($Q=4$) 
of Fig.~\ref{FIG:OpDepAPBC} since the $M_1 M_1$ correlator hardly overlaps
with the $n=0$ scattering state as shown in the right panel ($Q=3$) 
of Fig.~\ref{FIG:OpDepAPBC}.

\begin{figure}
\begin{center}
\includegraphics[angle=-90,scale=0.3]{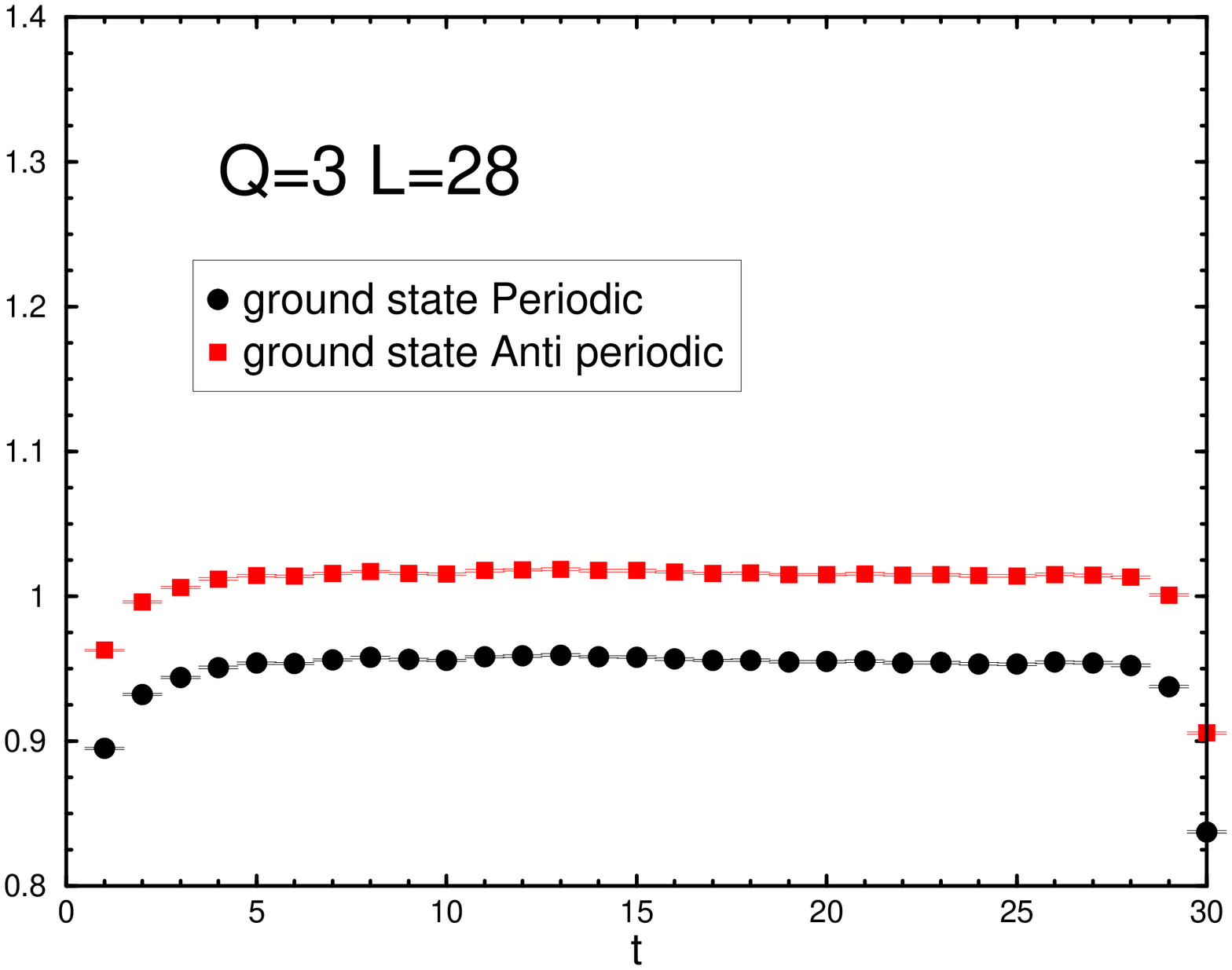}
\includegraphics[angle=-90,scale=0.3]{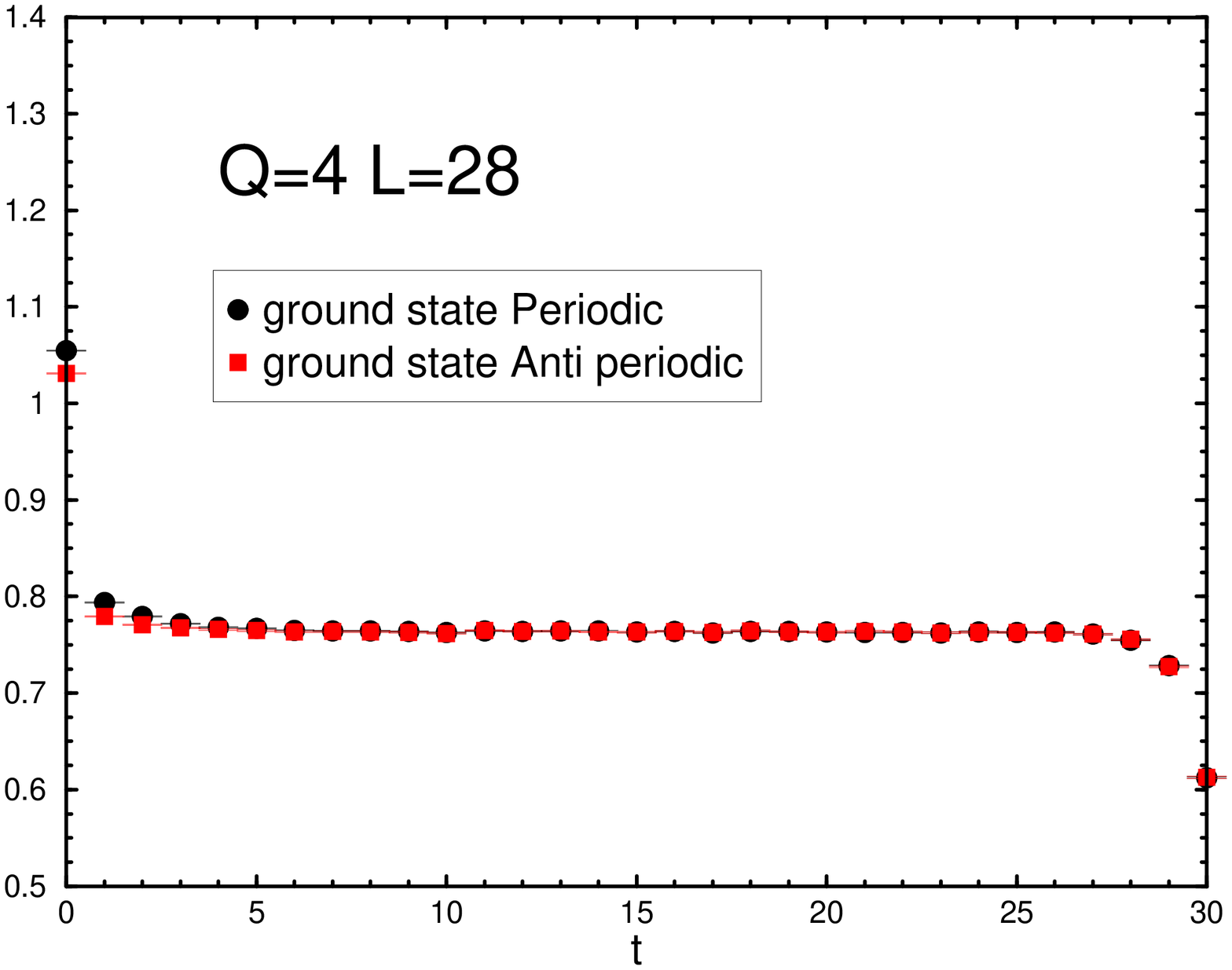}
\end{center}
\caption{The effective mass plots for the largest eigenvalue $\lambda_{\alpha}(t,t_0=7)$,
which corresponds to the lowest energy state, on the lattice with $L=28$
in the $^{1}S_0$ channel. 
The left (right) panel is for $Q=3$ ($Q=4$).
Full circles (full squares) represent  
the ground states from simulations with periodic (anti-periodic)
boundary conditions for electron fields in the spatial direction. 
}
\label{FIG:BcDepAPBC}
\end{figure}

We finally employ the diagonalization method to extract the ground states
in both $Q=3$ and $Q=4$ through the same procedure described in Sec.~\ref{Sec:4-B}.
Then we compare ground state energies for both $Q=3$ and $Q=4$
in the P.B.C. with those in the A.P.B.C. in Figs.~\ref{FIG:BcDepAPBC}. 
The left panel ($Q=3$), the effective mass of the ground state is clearly shifted 
up in changing from P.B.C. to A.P.B.C., while the plateau of the ground state 
doesn't change between P.B.C. and A.P.B.C. cases in the right panel ($Q=4$).
An energy shift in the $Q=3$ case is consistent with an estimation of 
$2(E^{0}_{e}({\rm A.P.B.C.})-E^{0}_{e}({\rm P.B.C.}))$.
We certainly confirm that the scattering state ($Q=3$) is sensitive to the spatial boundary
condition, while the bound state has no dependence of the spatial boundary condition for
the electron fields.

%
%
%

%
%

\clearpage
\section*{Appendix B: Volume dependence of the spectral weight}
\label{Apx-B}

The spectral decomposition of the matrix correlator is given by
%
%
\be
G_{ij}(t)=\sum_{\alpha=0} (v_\alpha)_{i} (v^{*}_\alpha)_{j} e^{-E_{\alpha}t}
\ee
with the spectral amplitude 
$(v_{\alpha})_{i} = {\langle 0 | \Omega_{i}|\alpha\rangle}_{V}$. 
A subscript in a ket ${|\alpha\rangle}_{V}$ stands for finite volume $V=L^3$. 
Remark that the finite-volume states ${|\alpha\rangle}_{V}$ are normalized to unity.
The spectral amplitudes are given by solving the following equations~\cite{Gockeler:1994rx}
%
%
\be
\sum_{i} (w_{\alpha})_{i}(v^{*}_{\beta})_{i}=\delta_{\alpha \beta} e^{E_{\alpha}t_0/2},
\ee
where $(w_{\alpha})_{i}$ is an $i$ component of vectors ${\bf w}_{\alpha}$,
which are determined through the following generalized eigenvalue problem~\cite{Gockeler:1994rx}:
\be
G(t){\bf w}_{\alpha}=\lambda_{\alpha}(t,t_0)G(t_0){\bf w}_{\alpha}.
\label{Eq:GEGVL}
\ee
To solve this eigenvalue equation, we have employed a diagonalization 
of the transfer matrix $M(t,t_0)=G^{-1/2}(t_0)G(t)G^{-1/2}(t_0)$, which
provides the same eigenvalues $\lambda_{\alpha}(t,t_0)=e^{-E_{\alpha}(t-t_0)}$ 
of Eq.(\ref{Eq:GEGVL})
%
%
\be
M(t,t_0){\bf u}_{\alpha}=\lambda_{\alpha}(t,t_0){\bf u}_{\alpha}
\ee
with the orthonormal eigenvectors ${\bf u}_{\alpha}=G^{1/2}(t_0){\bf w}_{\alpha}$, 
if $G(t)$ is an Hermite matrix~\cite{Gockeler:1994rx}.
The relative overlap between the chosen operator $\Omega_i$ and  
energy eigenstates ($\alpha=0,1,2,\cdot\cdot\cdot$) can be determined by 
the squared normalized amplitudes (the normalized spectral weights)
\be
(A_{\alpha})_{i}=\frac{|(v_{\alpha})_{i}|^2}{\sum_{\alpha}|(v_{\alpha})_{i}|^2}.
\ee
The normalized spectral weights calculated
in simulations at $L=28$ are tabulated in Table~\ref{Table:L28normAmpl}
as typical examples.

Here, we remind that the finite-volume states are normalized 
to unity, {\it regardless of whether the single particle state or the multi-particle
state}. Suppose the eigenstate $\alpha$ is a single particle state,
we simply obtain the correspondence between the finite-volume and the infinite-volume
states:
%
%
\be
{|\alpha\rangle}_{V}=\frac{1}{\sqrt{2E_{\alpha}V}}{|\alpha\rangle}_{\infty}
\;\;\;(\mbox{single particle state}),
\ee
where ${|\alpha\rangle}_{\infty}$ is normalized as $2E_{\alpha}$ particles
per unit volume.
On the other hand, if the eigenstate $\alpha$ is a two-particle state, 
the correspondence factor between ${|\alpha\rangle}_{\infty}$ and
${|\alpha\rangle}_{V}$ should depend on dynamics between
two particles. Such corrected factor is explicitly derived by Lellouch and
L\" uscher (denoted in the following by LL) to determine 
the physical $K\rightarrow \pi \pi$ amplitude
from the finite volume calculation~\cite{Lellouch:2000pv}.
Here, we consider the LL-factor only in the non-interacting case, where
the scattering phase shift between two particles is taken to be zero, 
for a simplicity, and then obtain the following correspondence between
${|\alpha\rangle}_{\infty}$ and
${|\alpha\rangle}_{V}$ for  
$S$-wave two-particle states~\cite{{Lellouch:2000pv},{Lin:2001ek}}:
%
%
\be
{|\alpha\rangle}_{V}\propto \frac{1}{E_{\alpha}V}{|\alpha\rangle}_{\infty}
\;\;\;
(\mbox{two-particle state}),
\label{Eq:TwoPState}
\ee
which indicates that observed spectral amplitudes
$(v_{\alpha})_{i}$ for the local-type operator ($i=P$) are proportional to 
$1/\sqrt{V}$ for the single particle state and proportional to $1/V$ for the $S$-wave 
two-particle state, since the physical spectral amplitude $\langle 0|\Omega_{P}{|\alpha\rangle}_{\infty}$ in the case of the local-type operator should 
not depend on the size of the spatial volume $V=L^3$. 

\begin{figure}[b]
\begin{center}
\includegraphics[angle=-90,scale=0.3]{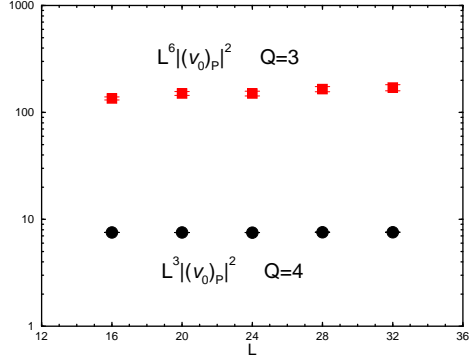}
\end{center}
\caption{
The volume dependence of the spectral weight of the $^1S_0$ ground state 
with the local-type operator $\Omega_P$.
We plot  $|\langle 0| \Omega_P | \alpha=0\rangle|^2$
scaled by $V^2=L^6$ for $Q=3$ (full squares) and by $V=L^3$ 
for $Q=4$ (full circles) as a function of spatial lattice size $L$.
Remind that $Q=3$ ($Q=4$) is the unbound system (the bound system),
where the $n=0$ scattering state (the bound state) is the ground state.
There is no appreciable $L$-dependence in either cases.  
}
\label{FIG:SpAmpl}
\end{figure}

Let us consider the volume dependence of the spectral amplitude 
of the ground state with the local-type operator $\Omega_P$.
In Fig.~\ref{FIG:SpAmpl}, we plot 
the finite-volume spectral weight $|\langle 0| \Omega_P {| \alpha=0\rangle}_V|^2$
scaled by $V^2$ for $Q=3$ and by $V$ 
for $Q=4$ as a function of spatial lattice size $L$.
Recall that $Q=3$ is the unbound system, while $Q=4$ is the bound system. 
No appreciable $L$-dependence is observed in either cases.  
This indicates that the finite-volume spectral weight for the local-type operator  
has a specific volume dependence according to whether the single particle
state or the two-particle state. In other words, each contribution from 
two-particle states (scattering states) relative to the single particle state 
is suppressed by a inverse of the volume factor, $1/L^3$ in the $PP$ correlator.

%
%

\clearpage

\end{document}